\documentclass[aps,prd]{revtex4-2}
\usepackage{color}
\usepackage{graphicx}
\usepackage{amsmath}
\usepackage{amssymb}
\usepackage{soul}
\usepackage[normalem]{ulem}
\usepackage[english]{babel}

\newcommand{\beq}{\begin{equation}}
\newcommand{\eeq}{\end{equation}}
\newcommand{\barr}{\begin{eqnarray}}
\newcommand{\earr}{\end{eqnarray}}

\newcommand{\mbf}{\mathbf}

\begin{document}

\title{Effective electromagnetic Lagrangians in the derivative expansion method}
\author{R. Mart\'\i nez von Dossow}
\email{ricardo.martinez@correo.nucleares.unam.mx}
\affiliation{Instituto de Ciencias Nucleares, Universidad Nacional Aut\'{o}noma de M\'{e}%
xico, 04510 Ciudad de M\'{e}xico, M\'{e}xico}
\author{Luis F. Urrutia}
\email{urrutia@nucleares.unam.mx}
\affiliation{Instituto de Ciencias Nucleares, Universidad Nacional Aut\'{o}noma de M\'{e}%
xico, 04510 Ciudad de M\'{e}xico, M\'{e}xico}

\begin{abstract}
We calculate the effective electromagnetic Lagrangian up to the lowest-order corrections in the derivatives for two fermionic systems of interest in condensed matter physics in the linearized approximation of the tight-binding Hamiltonian near the Fermi level in the Brillouin zone: (i) the $(3+1)$  description of the simplest Weyl semimetal and (ii) the massive $(2+1)$ electrodynamics,  which can serve as a model for the interface between two $(3+1)$ topological insulators. We employ the derivative expansion method which directly provides local effective Lagrangians and allows selecting from the outset both the powers of the electromagnetic potential to be considered together with the number of relevant derivatives. We find new higher-order derivative corrections to Carroll-Field-Jackiw electrodynamics. In general, the new terms we find either have a similar structure
or constitute a relativistic generalization of some recent phenomenological proposals found in the literature. In this way, they should be incorporated into these proposals for assessing the relative significance of all the terms included up to a given order.
\end{abstract}

\maketitle

%\today

\section{Introduction}

The rise of topological materials exhibiting a Dirac-like  behavior of their energy bands in regions close to the Fermi energy in the Brillouin zone has fostered a revival of many techniques already developed in
quantum field theory, in order to obtain their electromagnetic response codified in an effective action. The starting point is the condensed matter tight-binding Hamiltonian for the fermionic excitations, coupled to an external electromagnetic field, and linearized around the Fermi energy in the Brillouin zone.
This coupled Hamiltonian can be embedded in a Dirac-like theory similar to that described in the fermion-photon sector of the Standard Model Extension (SME) \cite{SME,SME2}, designed to study Lorentz and CPT violations in high-energy physics. The electromagnetic effective action is subsequently obtained by ``integrating'' the fermion fields, which yields corrections to the standard Maxwell's equations that encode the electromagnetic response of the material. 

Besides the intrinsic interest in effective electromagnetic Lagrangians,  additional motivation to investigate higher-order derivative contributions to them stems from recent publications introducing phenomenological additions of this kind to well-studied systems such as axion electrodynamics \cite{BARREDO}  and $2+1$  Maxwell-Chern-Simons  (MCS) electrodynamics \cite{TVM}, for example. The first reference considers the extension of electrodynamics to a medium with constitutive relations $\mbf{D}= \epsilon \mbf{E} +\psi(x) \boldsymbol{\nabla}^2 \mbf{B}$ and $\mbf{H}= \frac{1}{\mu}\mbf{B}-\psi(x) \boldsymbol{\nabla}^2 \mbf{E}$, dubbed $\psi$-electrodynamics, and studies its realization together with its possible applications in metamaterials. In reference \cite{TVM} an extra relativistic term proportional to $\xi \, \epsilon^{\mu\nu\rho} A_\mu \boldsymbol{\nabla}^2 F_{\nu\rho}$ is added to the MCS Lagrangian to account for the Hall viscosity $\xi$ of the system, which plays a fundamental role in elucidating the topological electromagnetic phases of matter. Originally, the viscosity parameter was introduced as a geometric response of the energy-momentum tensor arising from deformations in the underlying metric of the fluid \cite{AVRON}. Later it was shown that for Galilean invariant quantum Hall states the odd contribution to the viscosity tensor (yielding the  Hall viscosity)  appears in the electromagnetic response of the material as a momentum expansion in the conductivity \cite{HOYOS1,BRADLYN}. For a review, see, for example Ref. \cite{HOYOS2}. As anticipated, contributions of this type naturally emerge in higher-order derivative corrections within the framework of the effective action. This is what motivates our interest to investigate some of these contributions in order to compare them with those introduced in the cited references.

There is an ample variety of methods to deal with the determination of the electromagnetic response $S_{\rm eff}(A)$of fermionic systems,  and each of them has its own advantages and drawbacks, especially regarding the coordinate dependence of the external electromagnetic fields. 

For example, many calculations in condensed matter directly focus on the resulting current, which is obtained either via the Kubo linear response theory \cite{KUBO} or by a semiclassical Boltzmann approach including an anomalous velocity which takes care of the topological characteristics of the material \cite{BOLTZMANN}. These methods are perturbative and start directly from the electromagnetic fields, thus being gauge invariant {\it ab initio}. Nevertheless, they are mostly restricted to constant (slowly varying) or harmonic time-dependent fields.

The quantum field theory-motivated calculations yielding an effective action usually start from a Dirac-like action coupled to the electromagnetic field via the electromagnetic potentials. In this way, gauge invariance is not manifest, and it is mandatory to make it explicit at the end of the calculation.
These effective actions are known to all orders in the
electromagnetic fields only in a restricted family of backgrounds, such as constant fields or plane wave fields \cite{SCHWINGER},
to name some well-known cases. For a review, see, for
example, \cite{DUNE}. The extension to the case of nonhomogeneous electromagnetic fields has remained a subject of investigation and several advances have been reported \cite{11,12,13,14,15}. Also,  further developments on non-perturbative methods have received much attention recently. A detailed review of this topic, emphasizing advances from the last decade, is presented in Ref. \cite{26}. Among the new theoretical tools providing an alternative to standard Feynman diagram calculations, the semiclassical worldline instanton method, deriving from the ``first quantized''
approach to field theory, has proven particularly useful in
calculating effective actions and related quantities in
the study of QED processes in external fields \cite{27,28}.

In this work, we employ one of the many versions of the so-called derivative expansion (DE) method , which provides an alternative approach for calculating effective Lagrangians in the low-energy limit as an expansion in derivatives of the involved fields.
 Our focus is on the electromagnetic interaction of fermions $\Psi$ with the electromagnetic field $A_\mu$ through a generalized coupling
$e {\bar \Psi}\Gamma^\mu A_\mu \Psi$.  The early version of the DE technique to be discussed in this article was introduced in
Refs. \cite{Novikov84,Zuk85,Fraser85,Aitchison1985}.
Some previous works employing the same version of the  DE approach that we follow in this manuscript include:  (i) the study of higher-derivative corrections in Lorentz-violating massive QED theories, incorporating different parameters of the Standard Model extension \cite{Leite2013,Leite2012}, (ii) the work of Ref. \cite{Mariz2021} dealing with the inclusion of non-minimal terms appearing in a further extension of the SME \cite{Kostelecky:2018yfa} and (iii) the work of Ref. \cite{brito2008}, where the author analyzes the same Lorentz-breaking massless theory that we study in this manuscript, focusing on the generation of the Carroll-Field-Jackiw term using different regularization methods. Unlike our approach, in these references, the calculations are performed perturbatively in the Lorentz-violation parameter. Finally, in Ref. \cite{Anacleto:2014zoa}, higher-derivative corrections in $QED_{2+1}$ are calculated, but only for the axial part.

The effective action $S_{\rm eff}(A)$ yielding the electromagnetic response  is a sum of terms of the form
\beq
\frac{e^n}{n} {\rm Tr}\Big( S\Gamma^\mu A_\mu\Big)^n, \qquad n=1,2,\dots
\label{GENFORM}
\eeq
where $S$ is the  free fermion propagator (i.e., without the electromagnetic interaction) and ${\rm Tr}$ denotes the trace in coordinate as well as matrix space. Diagrammatically, this is represented by the sum of closed fermion loops ( since the fermions have been integrated out) with a number of $n$ photon external legs. The calculation of each loop yields a non-local expression which can be subsequently expanded in derivatives of the electromagnetic fields to produce local expressions, identified as contributions to the effective Lagrangian. The DE procedure
reverses this situation by allowing to select, from the very beginning, the number of fields $A_\mu$, as well as the number of derivatives one chooses to include in the desired contribution to the effective Lagrangian, yielding a local expression from the outset. Nevertheless, it goes without saying that increasing these numbers makes the calculation very involved. Also,  since we are adding different combinations of terms depending on $A_\mu$, usually in these cases it is a hard task to exhibit the required gauge invariant result. However, finding the correct combination provides a consistency check of the calculation.

The underlying concept in this version of the DE program is an alternative method for calculating  generalized  traces of products of momentum- and coordinate-dependent operators, as suggested by Eq. (\ref{GENFORM}). To be precise, we will 
deal with objects of the form
\begin{equation}
T^{(n)}\equiv \mathrm{Tr}\Big(M_1(\hat{p})C_1(\hat{x})\,M_2(\hat{p})C_2(\hat{%
x})\dots M_n(\hat{p})C_n(\hat{x})\Big),  \label{GENT}
\end{equation}
where $\hat{x}$ and $\hat{p}$ denote coordinate and momentum operators of
arbitrary dimension and metric. For the purpose of our applications we will mainly consider four dimensions, 
with the metric $\mathrm{diag}(+,-,-,-)$ such that 
\begin{equation}
[{\hat x}^\mu, {\hat p}_\nu]=-i\delta^\mu{}_\nu,  \label{CR}
\end{equation}
where ${\hat p}_\mu= i \partial_\mu$ in the coordinate representation. We
assume that each $\hat{p}$-dependent function in (\ref{GENT}) can be expanded in a power series
of the argument, while we deal with the exact expressions for the $\hat{x}$-dependent functions.

The basic idea is to use the commutation relations (\ref{CR}) to rearrange the expression (\ref{GENT}) in such a way that it can be written as a sum of terms where, by
convention, the momentum operators are written to the left and the
coordinate operators are written to the right 
\begin{equation}
T^{(n)}=\sum_N T^{(n)}_N \equiv \sum_N \mathrm{Tr} \Big(L^{(n)}_N({\hat p})
\, R^{(n)}_N({\hat x}) \Big).
\end{equation}
Once this is achieved, each contribution to the full trace can be
immediately calculated as the sum of 
\barr
T^{(n)}_N &=&\int d^4 x \langle x | L^{(n)}_N({\hat p}) \, R^{(n)}_N({\hat x}%
)|x \rangle= \int d^4 x \langle x | L^{(n)}_N({\hat p})|x \rangle \,
R^{(n)}_N({x}),\nonumber \\
&=& \int d^4 x \, \int \frac{d^4 p}{(2\pi)^4}\, L^{(n)}_N({\ p}%
) \, R^{(n)}_N({x}),
\label{TRACE}
\earr
in terms of the numerical functions $L_N(p)$ and $R_N(x)$. We are using the
scalar product 
\begin{equation} \label{braket}
\langle x| p \rangle=\frac{1}{\sqrt{(2 \pi)^4}} e^{-ip_\mu x^\mu}, \qquad
\langle p| x \rangle=\frac{1}{\sqrt{(2 \pi)^4}} e^{+ip_\mu x^\mu},
\end{equation}
together with the completeness of each orthonormal basis.

Before going into the details, we indicate the organization of the paper. In Section II, we explain the basic strategy for moving coordinate-dependent operators to the right and momentum-dependent operators to the left in expressions with the generic structure (\ref{GENT}). To simplify the notation, we use one-dimensional operators at this stage, but we also indicate the direct generalization to arbitrary dimensions, focusing on the four-dimensional case in particular.  Section III is divided into two parts. In the first subdivision, we provide a review of the calculation of the effective electromagnetic action $S_{\rm eff}(A)$ of a Dirac field as a power series in the electromagnetic potential. In the second part, by restricting to the contribution of second order in $A_\mu$, we compare the DE method  with the standard procedure for calculating this term. To show the equivalence, we express both alternatives in momentum space. Next, we present two applications where the DE technique is applied, emphasizing derivation of higher-order derivative corrections to the effective Lagrangians. In Section IV, we discuss $(3+1)$-dimensional massless electrodynamics with an additional coupling which is relevant to the effective description of Weyl semimetals in condensed matter. We recover Carroll-Field-Jackiw  electrodynamics together with its lowest-order derivative correction. The second application, in section V, deals with massive electrodynamics in $(2+1)$ dimensions, which can serve as a model for the interface between two topological insulators in $(3+1)$ dimensions. In this case, we calculate the effective action up to terms of fourth order in the derivatives. Finally, we close with section VI, which summarizes and discusses our results. Three appendices are included, containing the detailed evaluation of some integrals used throughout the manuscript. Our signature is $(+,-,-,-)$, we take $\epsilon^{0123}=+1$, and we follow the electrodynamics conventions of Ref. \cite{JACKSON}. We work in natural units $\hbar=c=1$.

\section{The Method}

\label{METHOD}

The basis of the method can be easily  explained in the one-dimensional case. Let
us consider the simplest situation 
\begin{equation}
T^{(2)}=\mathrm{Tr}\Big(M_1(\hat{p})C_1(\hat{x})\,M_2(\hat{p}) C_2(\hat{x})%
\Big),
\end{equation}
where we need to interchange $C_1 M_2$. Assuming $M_2$ has a power expansion in $\hat{p}$, we examine the contributions of ${\hat{p}}^n$ to the commutator $[C_1(\hat{x}), M_2(\hat{p})]$. We start from the basic commutator
\begin{eqnarray}
&&[C_1(\hat{x}), \hat{p}]=-i\frac{dC_1(\hat{x})}{d \hat{x}},
\earr
which we reorganize in the convenient form
\barr C_1(\hat{%
x})\,\hat{p}=\Big(%
\hat{p}-i\frac{d}{d \hat{x}} \Big) C_1(\hat{x}).  \label{P1} 
\earr
 To firmly establish the notation and conventions we calculate in detail the next contribution
\begin{eqnarray}
[C_1(\hat{x}), \hat{p}^2] &=& \hat{p} [C_1(\hat{x}), \hat{p}] + [C_1(\hat{x}%
), \hat{p}] \hat{p},   \notag \\
&=& 2\hat{p} [C_1(\hat{x}), \hat{p}]+[ [C_1(\hat{x}), \hat{p}], \hat{p}] 
\notag \\
&=& -2i \hat{p} \frac{dC_1(\hat{x})}{d \hat{x}}- \frac{d^2C_1(\hat{x})}{d 
\hat{x}^2}, 
\label{DETCOM2}
\end{eqnarray}
where we choose to write the $\hat{p}$ operators to the left and the $\hat{x}$ operators to the right.
Next we identify the term $C_1( {\hat x}) {\hat p}^2$ as 
\beq
C_1(\hat{x}) \hat{p}^2= \Big(\hat{p}^2-2i \hat{p} \frac{d}{d \hat{x}} -\frac{%
d^2}{d \hat{x}^2}\Big)C_1(\hat{x}).
\label{EXACTP2}
\end{equation}
This is an exact operator expression and Eq. (\ref{DETCOM2}) makes it clear that $\frac{d}{d \hat{x}}$ acts only in the operator $C_1$. The crucial step in the notation arises after rewriting  (\ref{EXACTP2}) as 
\begin{equation}
\Big(\hat{p}^2-2i \hat{p} \frac{d}{d \hat{x}} -\frac{d^2}{d \hat{x}^2}\Big)= %
\Big(\hat{p}-i \frac{d}{d \hat{x}}\Big)^2.
\end{equation}
In other words, from now on we are assuming that the operator $\hat{p}$  commutes with  $ \frac{d}{d \hat{x}}$,
which  naturally  act to the right upon the corresponding $C_1(\hat{x})$. That is to say, $ \frac{d}{d \hat{x}}$ will be considered a ${C}$-number in the following. Having established this convention we immediately obtain  the relations
\barr
C_1(\hat{x})\hat{p}^n= \Big(\hat{p}-i \frac{d}{d \hat{x}}\Big)^n 
\, C_1(\hat{x}),
\label{EXCHANGEN}
\earr  
obtained by multiplying the $(n-1)$ equation to the right by $\hat{p}$ and using (\ref{P1}). This yields the general result
\beq
C_1(\hat{x})M_2(\hat{p})=M_2\left(\hat{p}-i \frac{d}{d \hat{x}}\right) C_1(\hat{x}).
\label{GEN}
\eeq
 Here we recall  that the derivative $\frac{d}{d \hat{x}}$ acts only upon the exchanged operator $C_1$.  Let us emphasize that  the notation $M({\hat p}-i \frac{d}{d {\hat x}})$ indicates that the operator ${\hat p}$ has been shifted to ${\hat p} -i \frac{d}{d {\hat x}}$ in the power expansion of $M({\hat p})$. Equation (\ref{GEN}) is the key identity of the method, with the appropriate generalization to arbitrary dimensions.

Extending the above construction to more factors let us consider $T^{(3)}$, for example,
\beq
T^{(3)}= \mathrm{Tr}\Big(M_1(\hat{p})\Big[C_1(\hat{x})\,M_2(\hat{p}) C_2(\hat{x}) \,M_3(\hat{p})\Big] C_3(\hat{x})%
\Big),
\eeq
and focus on the term in square brackets
\begin{equation}
\Big( C_{1}(\hat{x})\,M_{2}(\hat{p})\Big) \Big( C_{2}(\hat{x})M_{3}(\hat{%
p})\Big) =M_{2}\left( \hat{p}-i\frac{d\;}{d\hat{x}_{1}}%
\right) \left[ C_{1}(\hat{x})\;M_{3}\left( \hat{p}-i\frac{d\;}{d\hat{x}_{2}}%
\right) \right] C_{2}(\hat{x}),
\label{3P}
\end{equation}%
where we use the prescription (\ref{GEN}) to move $C_1$ through $M_2$ and $C_2$ through $M_3$. We have labeled de derivatives as  $\frac{d}{d \hat{x}_i}$ to make sure that that they  only  act upon  the corresponding $C_i(\hat{x})$.    

We still have to move $C_{1}(\hat{x})\;\;$across $M_{3}(\hat{p})$ in order to have all $\hat{x}$-depending operators to the right. Again we start from the first  non-trivial  combination arising from the power expansion of $M_3$
\beq
C_1({\hat{x}})\left( \hat{p}-i\frac{d\;}{d\hat{x}_{2}}%
\right)^2=  \Big(C_1(\hat{x})\, \hat{p}^2-2i C_1(\hat{x})\,\hat{p} \frac{d}{d \hat{x}_2} -C_1(\hat{x})\,\frac{d^2}{d \hat{x}_2^2}\Big).
\eeq
Using the relations (\ref{EXCHANGEN}) in the first two terms and the $C$-number property of $\frac{d}{d \hat{x}_2}$ yields
\barr
C_1({\hat{x}})\left( \hat{p}-i\frac{d\;}{d\hat{x}_{2}}%
\right)^2 &=& \left( \left( \hat{p}-i\frac{d\;}{d\hat{x}_{1}}%
\right)^2 -2i \left( \hat{p}-i\frac{d\;}{d\hat{x}_{1}}%
\right) \frac{d\;}{d\hat{x}_{2}}-\frac{d^2}{d \hat{x}_2^2} \right) C_1({\hat{x}}), \nonumber \\
&=& \Big( \hat{p}-i\frac{d}{d\hat{x}_{1}} -i\frac{d\;}{d\hat{x}_{2}}  \Big)^2   C_1({\hat{x}}),
\earr
which enforces the $C$-number character of the derivatives $\frac{d}{d\hat{x}_{i}}$. The general expression results
\beq
C_1({\hat{x}})M_3\left( \hat{p}-i\frac{d\;}{d\hat{x}_{2}}%
\right)= M_3\left( \hat{p}-i\frac{d\;}{d\hat{x}_{2}}%
-i\frac{d\;}{d\hat{x}_{1}}\right)\, C_1({\hat{x}}),
\eeq
and the final result for the product in Eq. (\ref{3P}) is 
\beq
M_{2}\left( \hat{p}-i\frac{d\;}{d\hat{x}_{1}}%
\right)  M_3\left( \hat{p}-i\frac{d\;}{d\hat{x}_{2}}%
-i\frac{d\;}{d\hat{x}_{1}}\right)\, C_1({\hat{x}}) C_{2}(\hat{x}),
\eeq

The final result for $T$ $^{(3)}\;$is 
\begin{eqnarray}
T^{(3)}=\mathrm{Tr}\left( M_{1}(\hat{p})M_{2}\left( \hat{p}-i\frac{d\;}{d%
\hat{x}_{1}}\right) M_{3}\left( \hat{p}-i\frac{d\;}{d\hat{x}_{2}}-i\frac{d\;%
}{d\hat{x}_{1}}\right) C_{1}(\hat{x})C_{2}(\hat{x})C_{3}(\hat{x}).\right) 
\end{eqnarray}
Since the product of the operators $M_1$, $M_2$ and $M_3$ involve  only the operator $\hat{p}$ together with commuting objects $\frac{d\;}{d\hat{x}_{i}}$ and the operators $C_i(\hat{x})$ are all displaced to the right, $T^{(3)}$ has just the form indicated in Eq. (\ref{TRACE}) yielding the final value
\begin{equation}
T^{(3)}=\int dx \int \frac{dp}{(2\pi )}\left( M_{1}(p)M_{2}\left( p-i\frac{d\;}{%
dx_{1}}\right) M_{3}\left( p-i\frac{d\;}{dx_{2}}-i\frac{d\;}{dx_{1}}\right)
C_{1}(x)C_{2}(x)C_{3}(x)\right). 
\end{equation}

The procedure  is easily generalized to  products including more terms. Besides, the generalization to higher  dimensions is also direct. For example, extending $T^{(3)}$ to four dimensions yields
\begin{eqnarray}
T^{(3)} &=&\mathrm{Tr}\left[ M_{1}(\hat{p}_{\mu })C_{1}(\hat{x}^{\nu
})\,M_{2}(\hat{p}_{\mu })\left( C_{2}(\hat{x}^{\nu })M_{3}(\hat{p}_{\mu
})\right) C_{3}(\hat{x}^{\nu })\right]   \notag \\
T^{(3)} &=&\mathrm{Tr}\left( M_{1}(\hat{p}_{\mu })M_{2}\left( \hat{p}_{\nu
}-i\frac{\partial }{\partial \hat{x}_{1}^{\nu }}\right) M_{3}\left( \hat{p}%
_{\rho }-i\frac{\partial \;}{\partial \hat{x}_{2}^{\rho }}-i\frac{\partial \;%
}{\partial \hat{x}_{1}^{\rho }}\right) C_{1}(\hat{x})C_{2}(\hat{x}%
)C_{3}(\hat{x})\right), 
\end{eqnarray}%
where we have deleted the super-index labeling the coordinate operators and  functions. The final trace evaluation in coordinate space yields 
\begin{equation}\label{trace}
T^{(3)}=\int d^4x \Big[\int \frac{d^{4}p}{(2\pi )^4} \, \left( M_{1}(p_{\mu })M_{2}\left(
p_{\nu }-i\frac{\partial \;}{\partial x_{1}^{\nu }}\right) M_{3}\left(
p_{\rho }-i\frac{\partial }{\partial x_{2}^{\rho }}-i\frac{\partial }{%
\partial x_{1}^{\rho }}\right) \Big] C_{1}(x)C_{2}(x)C_{3}(x)\right). 
\end{equation}
Defining the operator 
\begin{equation}
\Pi\Big(\, i\frac{\partial}{\partial x_1^\nu }, \,\, 
i\frac{\partial}{\partial x_2^\rho }\Big)= 
\int \frac{d^{4}p}{(2\pi )^4} \,  M_{1}(p_{\mu })M_{2}\left(
p_{\nu }-i\frac{\partial \;}{\partial x_{1}^{\nu }}\right) M_{3}\left(
p_{\rho }-i\frac{\partial }{\partial x_{2}^{\rho }}-i\frac{\partial }{%
\partial x_{1}^{\rho }}\right),
\end{equation}
we present Eq. (\ref{trace}) in the compact form
\beq
T^{(3)}=\int d^4 x \, \Pi\Big(\, i\frac{\partial}{\partial x_1^\nu }, \,\, 
i\frac{\partial}{\partial x_2^\rho }\Big)\, C_{1}(x)C_{2}(x)C_{3}(x).
\label{T3PI}
\eeq

The final goal now is to determine the local function of $x$ inside the integral (\ref{T3PI}). The standard strategy is to expand the functions $M_i $ in powers of the derivatives times functions of the momentum, thus giving  the name of ``the derivative expansion'' to the method. The general form of the resulting expression is 
\beq
T^{(3)}=\int \frac{d^{4}p}{(2\pi )^4}d^{4}x\Big(N_0+ N^\mu_{1}(p) \partial_\mu + N^{\mu\nu}_2(p) \partial_\mu \partial_\nu + \cdots    \Big)  C_{1}(x)C_{2}(x)C_{3}(x), 
\eeq
where we must make sure that  each derivative  $\partial/\partial x^\nu_i$ acts on the corresponding function $C_i(x)$. The $p$-integration  can now be performed in the coefficient of each term, and the gradient operators end up  yielding a  local dependence on $C_i(x)$ and their  derivatives.

\section{The effective electromagnetic action}

\subsection{The general setup}

We illustrate the use of the DE method in the calculation of the electromagnetic response of Dirac fermions in different settings. The main advantage of the method is that one can select from the very beginning the type of effective action one is interested in by selecting the number of electromagnetic potentials involved together with the number of derivatives required.   

Let us start with the fermionic action in an external electromagnetic field $\;A_{\mu }$ 
\begin{equation}
S=\int d^{4}x\bar{\Psi}(\left( \Gamma ^{\mu }i\partial _{\mu }-M\right)
-e\Gamma ^{\mu }A_{\mu })\Psi , \quad {\hat p}_\mu= i\partial_\mu, \quad
A_\mu=A_\mu({\hat x}^\nu),  \label{ACTION1}
\end{equation}%
where $\Gamma^\mu$ and $M$ can be taken in the general form described by the fermionic sector of the Standard Model Extension \cite{SME,SME2}, for example. The action (\ref{ACTION1}) is frequently used as a benchmark in the field theory calculation of the electromagnetic response of Dirac-like materials in condensed matter physics .The resulting  effective electromagnetic action $S_{\rm eff}(A)$ is given by 
\begin{eqnarray}
\exp [iS_{\rm eff}(A)] &=&\int D\bar{\Psi}D\Psi \exp [i\int d^{4}x\bar{\Psi}%
(\left( \Gamma ^{\mu }{\hat p}_\mu-M\right) -e\Gamma ^{\mu }A_{\mu })\Psi ] ,
\notag \\
&=&\det (\left( \Gamma ^{\mu }{\hat p}_\mu-M\right) -e\Gamma ^{\mu }A_{\mu
}).
\end{eqnarray}
Let us introduce the non-interacting Green function  $ S({\hat p_\mu})={i}/{\left(
\Gamma^{\mu }{\hat p}_\mu-M\right) } $ and write 
\begin{equation}
(\left( \Gamma ^{\mu }{\hat p}_\mu-M\right) -e\Gamma ^{\mu }A_{\mu })=\left(
\Gamma ^{\mu } {\hat p}_\mu-M\right) \left[ 1-S({\hat p_\mu}) \, \Gamma ^{\alpha }\left(
-ieA_{\alpha }\right) \right] .
\end{equation}%
such that%
\begin{equation}
\det (\left( \Gamma ^{\mu }{\hat p}_\mu-M\right) -e\Gamma ^{\mu }A_{\mu
})=\det \left( \Gamma ^{\mu }{\hat p}_\mu-M\right) \det \left[ 1-S({\hat p_\mu})\, \Gamma
^{\alpha }\left( -ieA_{\alpha }\right) \right].
\end{equation}%
Here we discard the irrelevant normalization factor\ $\det \left( \Gamma
^{\mu }{\hat p}_\mu-M\right) $ and write 
\begin{equation}
iS_{\rm eff}(A)=\ln \det \left[ 1-S({\hat p_\mu})  \, \Gamma ^{\alpha }\left( -ieA_{\alpha
}\right) \right] ={\rm Tr}\ln \left[ 1- S({\hat p_\mu})  \, \Gamma ^{\alpha }\left( -ieA_{\alpha
}\right) \right],
\end{equation}%
after using the identity $\det M=\exp {\rm Tr} \ln M$. Here, ${\rm Tr}$ indicates the trace in coordinate space which includes the trace $\rm tr$ in the Dirac matrix space. The above expression
provides the basis for a perturbative calculation of $S_{eff}(A)$, starting
from the power expansion of the logarithm 
\begin{equation}
\ln (1-x)=\sum_{n=1}^{\infty }-\frac{1}{n}x^{n}.
\end{equation}
In this approach we initially  keep the operator-valued expressions of the propagator $S(\hat{p}_\mu)$ together with that of the vector potential $A^\mu(\hat{x}^\alpha)$ such that $[\hat{x}^\mu, \hat{p}_\nu]=-i \delta^\mu{}_\nu$, so that 
\begin{equation}
iS_{\rm eff}(A)=Tr\sum_{n=1}^{\infty }-\frac{1}{n} \Big( S({\hat p}_\mu) \,
\left( -ie\Gamma ^{\alpha }A_{\alpha }({\hat x}^\mu)\right) \Big)^{n}.
\label{ACTOP}
\end{equation}%
In the following we restrict ourselves to an expansion to second order in $A^\mu$, having up to four derivatives in the electromagnetic fields. The starting point is   
\begin{eqnarray}
\label{Seff1}
iS_{\rm eff}^{(2)}(A) &=&+\frac{e^{2}}{2}{\rm Tr}\left[ S({\hat p})\, \Gamma ^{\mu
}A_{\mu }({\hat x})\, S({\hat p})\, \Gamma ^{\nu }A_{\nu}({\hat x})\right].
\end{eqnarray}%

\subsection{Contact with the Feynman diagram calculation to one-loop}

In this subsection, we establish the equivalence between the following two alternative methods for calculating the one-loop effective electromagnetic action: (1) the use of the DE method which we follow in this work and (2) the usual Feynman diagram calculation  as shown in Ref. \cite{PESKIN}, for example.
On the one hand, the DE approach we have discussed naturally operates in coordinate space, yielding local results at each stage of the calculation. On the other hand, the standard calculation of $S_{\rm eff}^{(2)}(A)$  in coordinate space yields a non-local result. The point of contact is achieved going to the momentum representation in both cases,  as we now show.

Let us start with  the more familiar case of the Feynman diagram calculation, where we use the coordinate basis to evaluate (\ref{Seff1}) by taking the coordinate trace as $\int d^4 x \,   \langle x |\, i  
\,  S^{(2)}_{\rm eff } (A) |  x \rangle $, and subsequently inserting a unity as $\int d^4 x' |x'\rangle \langle x'|=I$ immediately before $\Gamma^\mu$.  Following this  method, we obtain the following  one-loop expression in coordinate space \cite{PESKIN}
\begin{equation}
iS_{\rm eff}^{(2)}(A)=\frac{e^{2}}{2}\int d^{4}x\;d^{4}x^{\prime }\;A_{\mu
}(x^{\prime }){\rm tr} \left[ S(x-x^{\prime }) \Gamma ^{\mu } S(x^{\prime }-x)
\Gamma ^{\nu }\right] A_{\nu }(x),
\label{ONELOOP}
\end{equation}
which is a non-local expression.
Here, $
\langle x|S({\hat p}) | x^{\prime}\rangle=S(x-x^{\prime}) $
is the free fermion propagator  (i.e. without the electromagnetic interaction)  denoted by $S(p)$ in momentum space. The only trace remaining is that in the Dirac matrix space. Notice that this way of taking the coordinate trace is a different operation, when compared to the prescription in the DE program discussed in section \ref{METHOD}, where we moved the coordinate-dependent operators to the right and the momentum-dependent operators to the left. 

To make contact with the DE approach we introduce the Fourier transform 
\begin{equation}
    A_\mu(x)=\int \frac{d^4k}{(2\pi)^4}e^{-ik_\mu x^\mu}A_\mu(k),
    \label{FT1}
\end{equation}
in Eq. (\ref{ONELOOP}), with $i\partial_\mu= k_\mu$ and the notation $k_\mu x^\mu=k \cdot x$,
obtaining 
\begin{equation}
iS_{\rm eff}^{(2)}(A) =\frac{e^{2}}{2}\int d^{4}x\;\frac{d^{4}k^{\prime }}{(2\pi )^{4}}\;\frac{%
d^{4}k}{(2\pi )^{4}}\;\Pi^{\mu\nu}(k')\,
e^{-ix\cdot(k^{\prime }+k)}A_{\mu }(k^{\prime })A_{\nu }(k).
\label{SMOM1}
\end{equation}
with  
\beq
\Pi^{\mu\nu}(k')=\int \frac{d^4p}{(2\pi)^4} {\rm tr}\Big( S( p)\, \Gamma ^{\mu
}\, S( p-k')\, \Gamma ^{\nu } \Big),
\label{VPTK}
\eeq
which is the standard form of vacuum polarization tensor. Usually the result (\ref{SMOM1}) is presented after the $d^4 x$ integration is performed, which in turns allows a further integration in one of the momenta. We keep the integration over $x$ because it is mandatory in the DE formulation as a consequence of its local character. 
In the next step of the calculation following Eq. (\ref{SMOM1}), one further combines the denominators of the propagators $S(p)$ and $S(p-k')$ to produce an exact result for $\Pi^{\mu\nu}(k')$. 

 Now let us consider the alternative DE  calculation which
 starts from the key identity  Eq. (\ref{GEN}), that allows to rewrite the effective action (\ref{Seff1}) as
\begin{equation}
    iS_{\rm eff}^{(2)}(A) =+\frac{e^{2}}{2}{\rm Tr}\left[ S({\hat p})\, \Gamma ^{\mu
}\, S({\hat p-i\partial'})\, \Gamma ^{\nu }A'_{\mu }({\hat x}) A_{\nu}({\hat x})\right],
\label{Seff2}
\end{equation}
where the superscript ``prime'' indicates where the derivative acts. Then, evaluating the trace as in (\ref{trace}), we obtain
\begin{equation}
    iS_{\rm eff}^{(2)}(A) =+\frac{e^{2}}{2}\int d^4x \, \int \frac{d^4p}{(2\pi)^4} {\rm tr}\Big( S( p)\, \Gamma ^{\mu
}\, S( p-i\partial')\, \Gamma ^{\nu } \Big) A'_{\mu }(x) A_{\nu}(x). 
\label{Seff3}
\end{equation}
Here  we recognize what we call the vacuum polarization operator, where $k'$ in Eq. (\ref{VPTK}) has been  replaced by $-i\partial'$, yielding
\beq
\Pi^{\mu\nu}(i\partial')=\int \frac{d^4p}{(2\pi)^4} {\rm tr}\Big( S( p)\, \Gamma ^{\mu
}\, S( p-i\partial')\, \Gamma ^{\nu } \Big),
\label{PIMUNUDER}
\eeq
together with 
\beq
iS_{\rm eff}^{(2)}(A) =+\frac{e^{2}}{2}\int d^4x \, \Big[\Pi^{\mu\nu}(i\partial') A'_{\mu }(x)\Big] A_{\nu}(x).
\label{SDER}
\eeq
As a final step to make the comparison, let us now rewrite (\ref{SDER}) in momentum space with
\beq
A'_{\mu }(x) A_{\nu}(x)= \int \frac{d^4k'}{(2\pi)^4}  \frac{d^4k}{(2\pi)^4} e^{-ik' \cdot  x} e^{-ik \cdot  x}A'_\mu(k') A_\nu(k).
\eeq
Since  $\partial'$ acts only in $A'_\mu(x)$ we can replace the operator $i\partial'$  by $k'$ in the vauum polarization operator (\ref{PIMUNUDER}) acting on $e^{-i k'\cdot x}$, 
such that the Fourier transformed   version of Eq. (\ref{SDER}) reads,
\beq
iS_{\rm eff}^{(2)}(A) =+\frac{e^{2}}{2}\int d^4x \, \frac{d^4k'}{(2\pi)^4}  \frac{d^4k}{(2\pi)^4} \, \Pi^{\mu\nu}(k') \,  e^{-ik' \cdot  x} e^{-ik \cdot  x}A_\mu(k') A_\nu(k),
\label{SEFFMOM1}
\eeq
in terms of the  momentum representation of the operator $\Pi^{\mu\nu}(k')$, given in Eq. (\ref{VPTK}), which arises in the Feynman diagram calculation.

Summarizing, we have shown that the full expression for the effective action $S_{\rm eff}^{(2)}(A)$  is the same as calculated by either of the two methods, as depicted in Eqs. (\ref{SMOM1}) and (\ref{SEFFMOM1}). In other words, the same expression in coordinate space  
can be achieved in terms of  an expansion in $k_\mu'$ of the Feynman diagram calculation of $\Pi^{\mu\nu}(k')$, after  going back to coordinate space taking care that the powers in $k'_\alpha$ would end up acting only as derivatives on $A'_\mu$ .

When the  main goal of a calculation  is to obtain a local expression for the effective action (\ref{ACTOP})  in coordinate space,  
we highlight the specific prescription for taking the trace in the DE method, described in section \ref{METHOD},  as an important advantage, particularly in the case when we are interested in higher derivative corrections to this effective action containing additional powers in $A_\mu$.

\section{First application: the efective action  in an extended  massless QED}
\label{IIIB}

In this section, we choose the additional coupling $b_\mu \gamma_5 \gamma^\mu$ to massless QED in $(3+1)$ dimensions to show the calculation of the effective action. We start from the fermionic action 
\begin{equation}	S=\int d^{4}x\bar{\Psi}(\left( \gamma ^{\mu }i\partial _{\mu }- b_\mu \gamma_5 \gamma^\mu \right)
	-e\gamma ^{\mu }A_{\mu })\Psi,
 \label{BMUACTION}
\end{equation}
where we identify  $\Gamma^\mu=\gamma^\mu$ and  $M=b_\mu \gamma_5 \gamma^\mu$ in the general  Eq. (\ref{ACTION1}). 
The additional coupling belongs to the fermionic sector of the SME and violates Lorentz and CPT symmetries. In condensed matter, the fermionic action describes the linearized approximation for the Hamiltonian of a Weyl semimetal with two cones, having energy separation $b_0$ and momentum separation $\mathbf{b}$ in the Brillouin zone. The general case, including tilting, anisotropy, and chemical potential without higher-order derivative corrections, is reported in Refs. \cite{URRU1, Gomez:2023jyl}.
Additional work  dealing with the use of fermionic actions like (\ref{ACTION1}), which violate Lorentz invariance,  to obtain the electromagnetic response of Dirac-like materials can be found in Refs. \cite{GRUSHIN,GRUSHIN1,ALANBABAK1,ALANBABAK2}.

As we will show next, the effective electromagnetic action resulting from  Eq. (\ref{BMUACTION}) describes Carroll-Field-Jackiw electrodynamics \cite{CFJ} plus higher-order derivative corrections. Since  $b_\mu$ is a constant vector, the effective action is given by the Eq. (\ref{ACTOP}) 
with the propagator 
\beq
S(\hat{p})=i/\left(\gamma^\mu \hat{p}_\mu-b_\mu \gamma_5 \gamma^\mu\right).
\label{NEWPROP} 
\eeq
where $M$ replaces the standard mass term.
We are interested in the effective action to second order in $A_\mu$, which is given by Eq. (\ref{Seff1}). In this case, Eqs. (\ref{PIMUNUDER}), (\ref{PIMUNUK}) and (\ref{SMOM1}) indicate that the result is given by the vacuum polarization tensor in any of its representations. Then, we focus on the calculation of this object. 

Using the relation (\ref{GEN}) and following the basic idea of the DE technique, we move $A_\mu({\hat{x}})$ to the right, obtaining Eq. (\ref{Seff2}).
    Next, we calculate the trace over the coordinate space as shown in Eq. (\ref{TRACE}),   arriving at Eq. (\ref{Seff3}),
which is written as
\begin{equation}
	S_{\rm eff}^{(2)}(A)=\frac{1}{2}\int d^4x \Pi^{\mu\nu} (\partial') A'_{\mu }(x)A_{\nu}(x), \qquad \Pi^{\mu\nu}(\partial')=ie^{2}{\rm tr}\int \frac{d^4p}{(2\pi)^4}\left[ S({ p})\, \gamma ^{\mu}\, S(p-i\partial')\, \gamma ^{\nu }\right],
\end{equation}
where $\Pi^{\mu\nu}(\partial')$
is the polarization tensor operator. 

Since we are dealing with a massless theory and the matrix $\gamma_5$ appears in the propagator (\ref{NEWPROP}),
we find it convenient to use chiral projectors.
We define the operators
\begin{equation}
	P_\chi=\frac{1+\chi\gamma_5}{2}, \hspace{5mm}\gamma_5^2=1, \hspace{5mm}P_++P_-=1, \hspace{5mm} P_\chi^2=1,\hspace{5mm}P_+P_-=P_-P_+=0,
\end{equation}
that project onto the right-handed (R) and left-handed (L) subspaces, with $\chi=+1$ and $\chi=-1$, respectively \cite{Altschul2004,Salvio2008,Scarpelli2008,Gomez:2023jyl}. Therefore, we can write the polarization tensor as the sum
 \begin{equation}
 	\Pi^{\mu\nu}=\Pi^{\mu\nu}_L+\Pi^{\mu\nu}_R,
 \end{equation}
with
\begin{equation}
	i\Pi^{\mu\nu}_\chi=e^{2}{\rm tr}\int \frac{d^4p}{(2\pi)^4}\left[ S({ p})\, \gamma ^{\mu}\, S(p-i\partial')\, \gamma ^{\nu }P_\chi\right].
\end{equation}
The action of the projectors on the propagator allows for the replacement of the matrix $\gamma^5$ in the denominator by the respective eigenvalues, yielding
\begin{equation}
	\frac{i}{\gamma^\mu p_\mu - b\gamma_5 \gamma ^\mu}\gamma^\nu P_\chi=P_\chi S_\chi(p)\gamma^{\mu},
\end{equation}
where
\begin{equation}
	S_\chi(p)=\frac{i}{\gamma^\mu p_\mu + \chi \gamma^\mu b_\mu}=\frac{i}{(p_\mu+\chi b_\mu)\gamma^\mu}=i\frac{(p+\chi b)_\mu \gamma^\mu}{(p+\chi b)^2}. 
\end{equation}
Using the cyclic property of the trace, we obtain
\begin{equation}
	i\Pi^{\mu\nu}_\chi=e^{2}{\rm tr}\int \frac{d^4p}{(2\pi)^4}\left[ S_\chi({ p})\, \gamma ^{\mu}\, S_\chi(p-i\partial')\, \gamma ^{\nu }P_\chi\right].
 \label{FINALPICHI}
\end{equation}

In the following, we consider only the axial contribution to the vacuum polarization, which amounts to retaining only the $\gamma^5$ piece of the projector $P_\chi$ in Eq. (\ref{FINALPICHI}). This restriction is motivated by some interesting results which emerge from this sector of the $\Pi^{\mu\nu}$ in condensed matter, such as the anomalous 
 Hall effect, together with contributions to the chiral magnetic effect \cite{CMHEFF}. 
Also, this sector allows for a simpler illustration of the DE method in connection with the derivation of gauge-invariant results. Then, the vacuum polarization operator reduces to 
\begin{equation}
	i\Pi^{\mu\nu}_\chi(\partial')=-\frac{\chi e^{2}}{2}{\rm tr}\int \frac{d^4p}{(2\pi)^4}\left[ \frac{(p+\chi b)_\alpha \gamma^\alpha}{(p+\chi b)^2}\, \gamma ^{\mu}\, \frac{(p-i\partial'+\chi b)_\beta \gamma^\beta}{(p-i\partial'+\chi b)^2}\, \gamma ^{\nu }\gamma^5\right].
\end{equation}
Defining $\pi_\alpha = (p+\chi b)_\alpha$, we have that
$(\pi-i\partial')^2=\pi^2-(2i\pi_\rho \partial'\rho + 
\partial^2)$, so we can rewrite 
\begin{equation} 
	i\Pi^{\mu\nu}_\chi(\partial')=-\frac{\chi e^{2}}{2}{\rm tr}\left(\gamma^\alpha\gamma^\mu\gamma^\beta\gamma^\nu\gamma^5\right)\int \frac{d^4p}{(2\pi)^4}\left[ \frac{\pi_\alpha}{\pi^2}\,  (\pi_\beta-i\partial'_\beta )\frac{1}{\pi^2-(2i\pi_\rho\partial'^\rho+\partial'^2)}\right],
 \label{PIMUNUDER2}
\end{equation}
where the trace is given by ${\rm tr}\left(\gamma^\alpha\gamma^\mu\gamma^\beta\gamma^\nu\gamma^5\right)= -4i\epsilon^{\alpha\mu\beta\nu}$ with $\epsilon^{0123}=+1$. The notation is $\pi^{2n}=(\pi_{\alpha} \pi^{\alpha})^n$.

Following the main objective of the DE technique, we now expand  $\Pi^{\mu\nu}_\chi(\partial') $ in powers of the derivatives $\partial'_\mu$. The last factor in Eq. (\ref{PIMUNUDER2}) yields
\begin{eqnarray}
	\frac{1}{\pi^2-(2i\pi_\rho\partial'^\rho+
 \partial'^2)}&=&\frac{1}{\pi^2}+\frac{1}{\pi^4}(2i\pi_\rho\partial'^\rho+
 \partial'^2)+\frac{1}{\pi^6}(2i\pi_\rho\partial'^\rho+
 \partial'^2)^2+\frac{1}{\pi^8}(2i\pi_\rho\partial'^\rho+
 \partial'^2)^3+..., \nonumber \\
 &\equiv& \mathcal{D}_0+\mathcal{D}_1+\mathcal{D}_2+\mathcal{D}_3+...,
\end{eqnarray}
where the lower index $i$ in $\mathcal{D}_i$ indicates the number of derivatives. Then we have
\begin{eqnarray}
\mathcal{D}_0=\frac{1}{\pi^2} \hspace{5mm}&,&\hspace{5mm}\mathcal{D}_1=\frac{2i\pi_\rho \partial'^\rho}{\pi^4}, \\
\mathcal{D}_2=\frac{\partial'^2}{\pi^4}-\frac{4\pi_\rho \pi_\eta\partial'^\rho\partial'^\eta}{\pi^6}\hspace{5mm}&,&\hspace{5mm}\mathcal{D}_3=\frac{2i\pi_\rho \partial'^\rho
{\partial'^2}}{\pi^6}-\frac{8i\pi_\rho \pi_\eta \pi_\gamma \partial'^\rho \partial'^\eta \partial'^\gamma}{\pi^8}.
\end{eqnarray}
Up to four derivatives, we expand
\begin{eqnarray}
		i\Pi^{\mu\nu}_\chi&=&2i\chi e^{2}\epsilon^{\alpha\mu\beta\nu}\int \frac{d^4p}{(2\pi)^4}\left[ \frac{\pi_\alpha}{\pi^2}\,  (\pi_\beta-i\partial'_\beta ) \left( \mathcal{D}_0+\mathcal{D}_1+\mathcal{D}_2+\mathcal{D}_3+...\right)\right],
\end{eqnarray}
which further simplifies to 
\begin{eqnarray}
		i\Pi^{\mu\nu}_\chi&=&-\chi e^{2}\epsilon^{\alpha\mu\beta\nu}\int \frac{d^4p}{(2\pi)^4}\left[ \frac{\pi_\alpha}{\pi^2}\,  \partial'_\beta  \left( \mathcal{D}_0+\mathcal{D}_1+\mathcal{D}_2+\mathcal{D}_3+...\right)\right],
  \label{piexpder}
\end{eqnarray}
due to the antisymmetry of the Levi-Civita tensor, which makes zero the term  proportional to $\pi_\alpha\pi_\beta$.

\subsection{The Carroll-Field-Jackiw (CFJ) contribution}

As a first application of the result (\ref{piexpder}), we show how  the Carroll-Field-Jackiw contribution to electrodynamics
\begin{equation}
\label{CFJ}
	S_{\rm CFJ}= C\, e^2 \, \int d^4x \,\epsilon^{\alpha\nu\beta\mu} \, b_\alpha A_\nu \partial_\beta A_\mu,
\end{equation}
arises from the effective action  corresponding to the fermionic coupling in Eq. (\ref{BMUACTION}). Here $C$ is a finite but undetermined numerical factor, as shown previously in the literature. See for example Refs. \cite{Jackiw200, Battistel2001,PerezVictoria, Andrianov2002, Alfaro2010, Altschu2019}.
 Table \ref{tab1} presents the different values of the constant $C$ obtained using various regularization methods in the calculations. For more details on the massive and massless cases, see \cite{petrovbook} and references therein.
	\begin{table}[h!]
    \centering
    \begin{tabular}{|c|c|c|}
        \hline
        \textbf{Method} & \textbf{Coefficient} \bf{$C$} & \textbf{Reference} \\ \hline
        Fujikawa                   & $-1/4\pi^2$        & \cite{Burkov2012} \\ \hline
        t'Hooft-Veltman regularization & $-1/4\pi^2$     & \cite{Assuncao2015} \\ \hline
        Cylindrical integration    & $-1/4\pi^2$      & \cite{Goswami2013,Gomez:2023jyl} \\ \hline
        Spherical integration      & $-1/16\pi^2$     & This work and \cite{Goswami2013} \\ \hline
        Dimensional regularization & $-1/8\pi^2$      & \cite{brito2008} \\ \hline
        Momentum cut-off regularization & $-1/8\pi^2$ & \cite{brito2008} \\ \hline
        Temperature regularization & $-1/8\pi^2$      & \cite{brito2008} \\ \hline
        Implicit regularization    & $-1/16\pi^2$     & \cite{Scarpelli2008} \\ \hline
    \end{tabular}
    \caption{Different values of the constant \( C \) according to the regularization method used in massless QED.}
    \label{tab1}
\end{table}
The first calculation of this term appeared in Ref. \cite{kostjack}, and since then it has been approached using various methods \cite{PerezVictoria,Scarpelli2008,brito2008,Burkov2012,Goswami2013,Assuncao2015}. 

To obtain  the CFJ contribution from  the DE approach, we must consider only the terms with one derivative of (\ref{piexpder}), i. e.
\begin{eqnarray}\label{CFJNI}
	i\Pi^{\mu\nu}_\chi&=&2i\chi e^{2}\epsilon^{\alpha\mu\beta\nu}\int \frac{d^4p}{(2\pi)^4}\left[\frac{\pi_\alpha \partial'_\beta}{\pi^2}\mathcal{D}_0\right] 
	=2\chi e^{2}\epsilon^{\alpha\mu\beta\nu}\int \frac{d^4p}{(2\pi)^4}\frac{\pi_\alpha}{\pi^4}\partial'_\beta.
\end{eqnarray}
Next, we need to solve the integral
\begin{equation}
\label{int1}
    I_\alpha=\int \frac{d^4p}{(2\pi)^4}\frac{\pi_\alpha}{\pi^4}=\int \frac{d^4p}{(2\pi)^4}\frac{(p+\chi b)_\alpha}{[(p+\chi b)^2]^2},
\end{equation}
which can be done in many ways, yielding finite but different results, which means that the integral is regularization-dependent.

One way to proceed is by using spherical coordinates, after performing a Wick rotation to Euclidean space, as shown in the Appendix \ref{spherical}. This yields
\begin{equation}
	i\Pi^{\mu\nu}_\chi (\partial')=\frac{i\chi^2 e^{2}}{16\pi^2}\epsilon^{\alpha\mu\beta\nu}b_\alpha \partial'_\beta.
\end{equation}
Reconstructing $\Pi^{\mu\nu}(\partial')$ and substituting in (\ref{SDER}), the effective action takes the form
\begin{equation}
	S^{(2)}_{\rm eff}(A)=\frac{ e^{2}}{16\pi^2}\epsilon^{\alpha\nu\beta\mu}\int d^4x b_\alpha  A_\nu \partial_\beta A_\mu,
\end{equation}
which coincides with (\ref{CFJ}) with the constant $C=-\frac{1}{16\pi^2}$. 
If we compute the integral (\ref{int1}) in cylindrical coordinates, we obtain a different value for $C$, given by $C = -\frac{1}{4\pi^2}$  \cite{Gomez:2023jyl}. On the other hand, if we use a naive shift $p_\alpha+\chi b_\alpha \rightarrow p_\alpha$ to evaluate the integral (\ref{int1}),  we get $C=0$ \cite{Goswami2013}.

\subsection{Higher-derivative  contribution in Carroll-Field-Jackiw electrodynamics}
The leading order high-derivative term is given by 
\begin{equation}\label{HDCFJ}
	S_{{\rm HD-CFJ}}(A)=  C'\, e^2 \, \int d^4x  \, \epsilon^{\alpha\nu\beta\mu}\, b_\alpha A_\nu 
 \, \partial^2 \,  \partial_\beta A_\mu,
\end{equation}
where  $C'$ is a finite but undetermined numerical factor, similarly to the case of the CFJ term. 

The rise of higher-order derivative terms deriving from the effective action has been studied in the literature for massive QED, starting from different contributions in the fermionic sector of the SME, corresponding to parameters like $b_\mu$ \cite{Leite2013} and $g_{\alpha\beta\gamma}$ \cite{Leite2012,Mariz2011}, for example. Also, non-minimal terms \cite{Mariz2021} and combinations of $b_\mu$ and non-minimal terms \cite{Mariz2011.2} have been considered. Recently, $S_{{\rm HD-CFJ}} $ has been proposed as part of the effective action for certain types of metamaterials \cite{barredo}. 

In this work, we consider, for the first time, the emergence of $S_{{\rm HD-CFJ}}$ from  a massless fermionic theory.
To generate this term, we need three derivatives acting on the field $A_\mu$. Therefore, we next  consider only such contributions in Eq. (\ref{piexpder}), i. e. 
\begin{eqnarray}\label{CFJHDNI}
	i\Pi^{\mu\nu}_\chi&=&2i\chi e^{2}\epsilon^{\alpha\mu\beta\nu}\int \frac{d^4p}{(2\pi)^4}\left[\frac{\pi_\alpha \partial'_\beta}{\pi^2}\mathcal{D}_2\right] 
 =2\chi e^{2}\epsilon^{\alpha\mu\beta\nu}\int \frac{d^4p}{(2\pi)^4}\left(\frac{\pi_\alpha}{\pi^6}\partial'_\beta 
 \partial'^2-4\frac{\pi_\alpha \pi_\rho\pi_\eta}{\pi^8}\partial'_\beta \partial'^\rho\partial'^\eta \right).
\end{eqnarray}
Solving these integrals in spherical coordinates (see Appendix \ref{spherical}), we obtain
\begin{equation}
	i\Pi^{\mu\nu}_\chi=i\frac{\chi^2 e^{2}}{48\pi^2b^2}\epsilon^{\alpha\mu\beta\nu}\, b_\alpha A_\nu \, \partial'^2 \, \partial_\beta A'_\mu.
\end{equation}
Therefore, the effective action is given by
\begin{equation}
		S_{\rm eff}^{(2)}(A)= -\frac{e^2}{48\pi^2b^2}\int d^4x \,  \epsilon^{\alpha\nu\beta\mu}\, b_\alpha A_\nu 
  \, \partial^2 \, \partial_\beta A_\mu.
\end{equation}
which coincides with the expression (\ref{HDCFJ}), choosing  $C'=-\frac{1}{48\pi^2b^2}$. The calculation in cylindrical coordinates yields zero, showing  no generation of a higher-derivative CFJ term of order two.

\section{Second application: the effective action  in $QED_{2+1}$ with higher order derivative corrections }
\label{IIIC}
In this case, we consider massive quantum electrodynamics in $(2+1)$ dimensions, which can serve as a model to describe the interface between two topological insulators in $(3+1)$ dimensions, as already mentioned in the Introduction. 
We start from the action 
\begin{equation}
    S=\int d^3 x \, \bar{\psi}\left(\gamma^\mu i\partial_\mu+e\gamma^\mu A_\mu-m\right) \psi,
\end{equation}
which again yields the effective electromagnetic action
\begin{equation}
	iS_{\rm eff}^{(2)}(A)=+\frac{e^2}{2}\mathrm{Tr}\left(S(\hat{p})\gamma^\mu A_\mu(\hat{x})S(\hat{p})\gamma^\nu A_\nu(\hat{x})\right),
\end{equation}
up to the first non-trivial order in an expansion in powers of $A_\mu$. 
Now,
\beq
S(p)=i/(\gamma^\mu p_\mu-m), \label{DIRACPROP}
\eeq
is the usual free fermionic propagator. 
Again, using the relation (\ref{GEN}), we move all momentum operators to the left, as in Eq. (\ref{Seff2}), and then we can easily perform the trace,  obtaining Eq. (\ref{Seff3}) where $
S(p)$ is now given by (\ref{DIRACPROP}). We write this last expression in momentum space as 

\begin{equation}
    iS_{\rm eff}^{(2)}(A)=\frac{e^2}{2}\mathrm{tr}\int\frac{d^3k'}{(2\pi)^3} \int \frac{d^3p}{(2\pi)^3} S(p)\gamma^\mu S(p-k')\gamma^\nu A'_\mu(k')A_\nu(-k'),
\end{equation}
where we recall that the momentum $k'_\mu$ acts as the derivative $\partial_\mu$ only in the field $A'_\mu(x)$ when going back to coordinate space. The polarization tensor takes the form
\begin{eqnarray}
    i\Pi^{\mu\nu}(k')&=&-e^2\mathrm{tr}\int\frac{d^3p}{(2\pi)^3} \left(\frac{\left(\gamma^\alpha p_\alpha+m\right)\gamma^\mu\left(\gamma^\beta p_\beta -\gamma^\beta k'_\beta+m\right)\gamma^\nu}{\left(p^2-m^2\right)\left((p-k')^2-m^2\right)}\right), \nonumber \\
	&\equiv&-e^2\int\frac{d^3p}{(2\pi)^3} \left(\frac{\mathrm{tr}\mathcal{N}^{\mu\nu}}{\left(p^2-m^2\right)\left((p-k')^2-m^2\right)}\right)  .
    \label{PI3+1}
\end{eqnarray}
Calculating the trace, we have
\begin{eqnarray}
\mathrm{tr}\mathcal{N}^{\mu\nu}&=&\mathrm{tr}\left[\left(\gamma^\alpha p_\alpha+m\right)\gamma^\mu\left(\gamma^\beta p_\beta -\gamma^\beta k'_\beta+m\right)\gamma^\nu\right], \nonumber \\
	&=&4p^\mu p^\nu -2(p^2-m^2)\eta^{\mu\nu}-2p^\mu k'^\nu+2 p_\alpha k'^\alpha \eta^{\mu\nu}-2p^\nu k'^\mu+2im\epsilon^{\mu\beta\nu} k'_\beta, \nonumber \\
	&=& \mathcal{N}^{\mu\nu}_0+\mathcal{N}^{\mu\nu}_1(k').
\end{eqnarray}
Here, we have split ${\cal N}^{\mu\nu}$ into the terms $N^{\mu\nu}_i$, where $i$ indicates the number of external momenta $k'_\mu$ contributing to the higher-order derivative terms through $\partial'_\mu$. The explicit expressions are 
\begin{equation}
	\mathcal{N}_0=4p^\mu p^\nu -2(p^2-m^2)\eta^{\mu\nu},
\end{equation}
\begin{equation}
	\mathcal{N}_1(k')=-2p^\mu k'^\nu+2p_\alpha k'^\alpha \eta^{\mu\nu}-2p^\nu k'^\mu+2im\epsilon^{\mu\beta\nu}k'_\beta.
\end{equation}
Now, we perform an expansion of the denominator in (\ref{PI3+1}) in terms of the momentum $k'_\alpha$, up to terms of order four
\begin{eqnarray}
	\frac{1}{(p-k')^2-m^2}\
	&=&  \frac{1}{p^2-m^2}+\frac{1}{\left(p^2-m^2\right)^2}(2p_\lambda k'^\lambda-k'^2)+\frac{1}{\left(p^2-m^2\right)^3}(2p_\lambda k'^\lambda-k'^2)^2\nonumber \\ 
	&&+\frac{1}{\left(p^2-m^2\right)^4}(2p_\lambda k'^\lambda-k'^2)^3+\frac{1}{\left(p^2-m^2\right)^5}(2p_\lambda k'^\lambda+k'^2)^4+..., \nonumber \\	&\equiv&\mathcal{D}_0+\mathcal{D}_1+\mathcal{D}_2+\mathcal{D}_3+\mathcal{D}_4.
\end{eqnarray}
We have separated the above expression as a sum of terms $D_i$, where $i$ indicates the number of  external momenta in each term. Collecting the terms yields
\begin{eqnarray}
\mathcal{D}_0=\frac{1}{p^2-m^2} \hspace{5mm}&,&\hspace{5mm}\mathcal{D}_1=\frac{2p_\lambda k'^\lambda}{\left(p^2-m^2\right)^2}, \\
\mathcal{D}_2=-\frac{1}{\left(p^2-m^2\right)^2}k'^2+\frac{4p_\lambda p_\sigma}{\left(p^2-m^2\right)^3}k'^\lambda k'^\sigma \hspace{5mm}&,&\hspace{5mm}\mathcal{D}_3=-\frac{4p_\lambda}{\left(p^2-m^2\right)^3}k'^2 k'^\lambda+\frac{8p_\lambda p_\sigma p_\eta}{\left(p^2-m^2\right)^4}k'^\lambda k'^\sigma k'^\eta,
\end{eqnarray}
and
\begin{equation}
	\mathcal{D}_4=\frac{k'^4}{(p^2-m^2)^3}-12\frac{p_\lambda p_\kappa}{(p^2-m^2)^4}k'^2 k'^\lambda k'^\kappa +16\frac{p_\lambda p_\kappa p_\sigma p_\eta}{(p^2-m^2)^5}k'^\lambda k'^\kappa k'^\sigma k'^\eta.
\end{equation}
Therefore, the vacuum polarization tensor takes the form
\begin{equation}
\Pi^{\mu\nu}=ie^2\int\frac{d^3p}{(2\pi)^3} \frac{\mathcal{N}_0+\mathcal{N}_1}{\left(p^2-m^2\right)}  \left(\mathcal{D}_0+\mathcal{D}_1+\mathcal{D}_2+\mathcal{D}_3+\mathcal{D}_4+...\right) ,
\end{equation}
where it is easy to identify each term by the number of external momenta, $k'_\mu$. Next, we will calculate all the terms in the polarization tensor that contain up to 4 external momenta. i.e.,
\begin{equation}
	\Pi^{\mu\nu}=\Pi^{\mu\nu}_{0k'}+\Pi^{\mu\nu}_{1k'}+\Pi^{\mu\nu}_{2k'}+\Pi^{\mu\nu}_{3k'}+\Pi^{\mu\nu}_{4k'}.
\end{equation}
Substituting the above expression in Eq. (\ref{SDER}), adapted to $2+1$ dimensions,
yields an effective action with higher-order derivative corrections up to order four. Next we calculate each separate contribution to $\Pi^{\mu\nu}$.

\subsection{Terms without external momentum}
They are included in
\begin{eqnarray}
	\Pi^{\mu\nu}_{0k'}
	&=&ie^2\int\frac{d^3p}{(2\pi)^3} \frac{\mathcal{N}_0 \mathcal{D}_0}{\left(p^2-m^2\right)}, 
 \label{PI0K}
\end{eqnarray}
which has the explicit  expression
\begin{eqnarray}
	\Pi^{\mu\nu}_{0k'}
	&=&ie^2\int\frac{d^3p}{(2\pi)^3} \left(\frac{4p^\mu p^\nu-2(p^2-m^2)\eta^{\mu\nu}}{\left(p^2-m^2\right)^2} \right)  
	=ie^2 \eta^{\mu\nu}\left(\frac{4}{3}\int\frac{d^3p}{(2\pi)^3} \frac{p^2}{\left(p^2-m^2\right)^2}-2\int\frac{d^3p}{(2\pi)^3} \frac{1}{\left(p^2-m^2\right)} \right), \nonumber \\ 
	&\equiv& ie^2 \eta^{\mu\nu}\left(\frac{4}{3}I^1_2-2I^0_1 \right).
 \label{PI0K1}
\end{eqnarray}
In reducing $\Pi^{\mu\nu}_{0k'}$, we have used the identity (\ref{Ipp}). The last line in Eq. (\ref{PI0K1}) is presented in terms of the master integral $I_\alpha^\beta$, defined in Eq. (\ref{IM}), and calculated in Appendix \ref{APMI}. The result is  
\begin{eqnarray}
	\Pi^{\mu\nu}_{0k}
	&=& ie^2 \eta^{\mu\nu}\left(\frac{4}{3}\frac{3i|m|}{8\pi}-2\frac{i|m|}{4\pi} \right)=0.  
\end{eqnarray}
In other words, there is no contribution without external momentum to the polarization tensor. This is expected to ensure gauge invariance.

\subsection{Terms with one external momentum}
The contributions are summarized in
\begin{eqnarray}
	\Pi^{\mu\nu}_{1k'}&=&ie^2 \int\frac{d^3p}{(2\pi)^3} \frac{\mathcal{N}_0 \mathcal{D}_1+\mathcal{N}_1\mathcal{D}_0}{\left(p^2-m^2\right)}   \nonumber \\
	&=& ie^2\int \frac{d^3p}{(2\pi)^3}  \frac{1}{\left(p^2-m^2\right)}\left((4p^\mu p^\nu -2(p^2-m^2)\eta^{\mu\nu}) \frac{2p_\lambda k'^\lambda}{\left(p^2-m^2\right)^2} \right. \nonumber \\ && \left. +\left(-2p^\mu k'^\nu+2p_\alpha k'^\alpha \eta^{\mu\nu}-2p^\nu k'^\mu \right) \frac{1}{p^2-m^2}+2im \, \epsilon^{\mu\beta\nu} \, k'_\beta\frac{1}{p^2-m^2}. \right) 
\end{eqnarray}
Note that only the term proportional to the Levi-Civita tensor survives, because integrals with an odd number $p_\mu$  cancels due to symmetric integration, as shown  in  Eq.  (\ref{Ip}). Therefore, we end up with
\begin{eqnarray}\label{pi1k}
	\Pi^{\mu\nu}_{1k'}&=& -2me^2 \int \frac{d^3p}{(2\pi)^3}  \frac{1}{\left(p^2-m^2\right)^2}\, \epsilon^{\mu\beta\nu}\, k'_\beta = -2me^2\, I^0_2 \,  \epsilon^{\mu\beta\nu} \, k'_\beta.
\end{eqnarray}
Using the master integral (\ref{IM}), we obtain
\begin{equation}
		\Pi^{\mu\nu}_{1k'}=-i\frac{e^2}{4\pi} \,  \mathrm{sgn}(m) \, \epsilon^{\mu\beta\nu} \, k'_\beta,  
\end{equation}
where $\mathrm{sgn}(m)$ is the sign function of $m$. This is the usual Chern-Simons term.
\subsection{Terms with two external momenta}
These contributions arise from
\begin{eqnarray}
	\Pi^{\mu\nu}_{2k'}
	&=&ie^2\int\frac{d^3p}{(2\pi)^3} \frac{1}{\left(p^2-m^2\right)}  \left(\mathcal{N}_0\mathcal{D}_2 +\mathcal{N}_1\mathcal{D}_1\right) \nonumber \\
	&=&ie^2 \int\frac{d^3p}{(2\pi)^3} \frac{1}{\left(p^2-m^2\right)}  \left(\left(4p^\mu p^\nu -2(p^2-m^2)\eta^{\mu\nu}\right)\left(-\frac{1}{\left(p^2-m^2\right)^2}{ k'}^2 +\frac{4p_\lambda p_\sigma}{\left(p^2-m^2\right)^3}{k'}^\lambda k'^\sigma\right)\right. \nonumber \\ && \left.  +\left(-2p^\mu k'^\nu+2p_\alpha k'^\alpha \eta^{\mu\nu}-2p^\nu {k'}^\mu\right)\left(\frac{2p_\lambda {k'} ^\lambda}{\left(p^2-m^2\right)^2}\right)+2im´\, \epsilon^{\mu\beta\nu} \, k'_\beta\left(\frac{2p_\lambda {k'}^\lambda}{(p^2-m^2)^2}\right)   \right).  \end{eqnarray}
Disregarding all odd terms in $p_\mu$, using the identities (\ref{Ipp}) and (\ref{Ipppp}), and organizing the terms, we obtain
\begin{eqnarray}
\Pi^{\mu\nu}_{2k'} &=& -ie^2  \left(\frac{8}{3}I^1_3-\frac{16}{15}I^2_4-2I^0_2\right)\eta^{\mu\nu} k'^2+\left(\frac{8}{3}I^1_3-\frac{32}{15}I^2_4\right) k'^\mu k'^\nu.  
\end{eqnarray}
In terms of the master integral (\ref{IM}), we arrive  at 
\begin{equation}
 \Pi^{\mu\nu}_{2k'}=-\frac{e^2}{12\pi|m|} \left(\eta^{\mu\nu}k'^2 -k'^\mu k'^\nu\right). 
 \label{PIMUNUK2}
\end{equation}
Note that this term is explicitly gauge invariant. We consider a highly non-trivial consistency check of the calculation: the fact that all five integrals, with their weird coefficients in Eq.(\ref{PIMUNUK2}) manages to produce a manifestly gauge-invariant result. In fact, the difficulty in making gauge invariance manifest is one of the drawbacks of the method when applied to  more complicated scenarios. Since we do not start from an explicitly gauge-invariant  expression, the different independent contributions provided by  the DE prodedure may complicate their necessary  rearrangement, finally yielding a manifestly gauge-invariant result. A possible way to alleviate this complication could be to use the Fock-Schwinger gauge for $A_\mu$ \cite{FSG}, as has been shown in different contexts in Refs. \cite{USEOFFSG}. 
\subsection{Terms with three external momenta}
They are included in
\begin{eqnarray}
	\Pi^{\mu\nu}_{3k'}
	&=&ie^2\int\frac{d^3p}{(2\pi)^3} \frac{1}{\left(p^2-m^2\right)}  \left(\mathcal{N}_0\mathcal{D}_3 +\mathcal{N}_1\mathcal{D}_2\right)  A_\nu \nonumber \\
&=&ie^2\int\frac{d^3p}{(2\pi)^3} \left(\frac{4p^\mu p^\nu -2(p^2-m^2)\eta^{\mu\nu}}{\left(p^2-m^2\right)}\left(\frac{-4p_\lambda}{\left(p^2-m^2\right)^3}k'^\lambda k'^2+\frac{8p_\lambda p_\sigma p_\eta}{\left(p^2-m^2\right)^4}k'^\lambda k'^\sigma k'^\eta \right)\right.\nonumber\\
&&\left.-\frac{2im\epsilon^{\mu\beta\nu}k'_\beta}{\left(p^2-m^2\right)}\left(\frac{1}{\left(p^2-m^2\right)^2}k'^2-\frac{4p_\lambda p_\sigma}{\left(p^2-m^2\right)^3}k'^\lambda k'^\sigma\right) \right).
\end{eqnarray}
Disregarding the odd terms in $p_\mu$, we have
\begin{equation}\label{pi3k}
\Pi^{\mu\nu}_{3k'}=	2me^2\epsilon^{\mu\beta\nu}k'_\beta\int\frac{d^3p}{(2\pi)^3} \left(\frac{1}{\left(p^2-m^2\right)^3}k'^2-\frac{4p_\lambda p_\sigma}{\left(p^2-m^2\right)^4}k'^\lambda k'^\sigma\right).
\end{equation}
Now, using the identity (\ref{Ipp}) and reorganizing terms, we arrive at
\begin{equation}
	\Pi^{\mu\nu}_{3k'}=2e^2m\, \epsilon^{\mu\beta\nu} \left(I^0_3-\frac{4}{3}I^1_4\right) k'^2 k'_\beta.
\end{equation}
Using the master integral (\ref{IM}), we finally obtain
\begin{equation}\label{SHD}
	\Pi^{\mu\nu}_{3k'} =	-\frac{ie^2}{48\pi}\frac{\mathrm{Sgn}(m)}{m^2} \, \epsilon^{\mu\beta\nu} k'^2 k'_\beta. 
\end{equation}
This term is the lowest-order higher-derivative counterpart of the Chern-Simons term.
\subsection{Terms with four external momenta}
Finally, we consider the contribution in  
\begin{eqnarray}
	\Pi^{\mu\nu}_{4k'}
	&=&ie^2\int\frac{d^3p}{(2\pi)^3} \frac{1}{\left(p^2-m^2\right)}  \left(\mathcal{N}_0\mathcal{D}_4 +\mathcal{N}_1\mathcal{D}_3\right) \nonumber \\
	&=&ie^2\int\frac{d^3p}{(2\pi)^3} \left((4p^\mu p^\nu -2(p^2-m^2)\eta^{\mu\nu})\left(\frac{ k'^4 }{(p^2-m^2)^4}-12\frac{p_\lambda p_\kappa}{(p^2-m^2)^5}k'^\lambda k'^\kappa k'^2+16\frac{p_\lambda p_\kappa p_\sigma p_\eta}{(p^2-m^2)^6}k'^\lambda k'^\kappa k'^\sigma k'^\eta\right) \right. \nonumber \\ && 
 \left. +(2p^\mu k'^\nu-2p^\alpha k'_\alpha \eta^{\mu\nu}+ 2p^\nu k'^\mu-2im \, \epsilon^{\mu\beta\nu} \, k'_\beta)\left(\frac{4p_\lambda}{\left(p^2-m^2\right)^4}k'^\lambda k'^2-\frac{8p_\lambda p_\sigma p_\eta}{\left(p^2-m^2\right)^5}k'^\lambda  k'^\sigma k'^\eta\right) \right). 
\end{eqnarray}
Disregarding the odd terms in $p_\mu$, using identities (\ref{Ipp}), (\ref{Ipppp}) and (\ref{I6p}), and  reorganizing terms, we arrive at 
\begin{eqnarray}
	\Pi^{\mu\nu}_{4k'}
	&=&ie^2 \left[  \left(\frac{20}{3}I^1_4+\frac{192}{105}I^3_6-2I^0_3-\frac{96}{15}I^2_5\right) \eta^{\mu\nu}k'^4 
	+\left( \frac{768}{105}I^3_6-\frac{192}{15}I^2_5+\frac{16}{3}I^1_4\right) k'^\mu k'^\nu k'^2   \right].
    \label{PI2+1k4}
\end{eqnarray}
Finally, using the master integral (\ref{IM}), and, in spite of the complicated structure of the coefficients in Eq. (\ref{PI2+1k4}) we obtain
\begin{equation}
	\Pi^{\mu\nu}_{4k'}=\frac{e^2}{120\pi|m|^3} \left[k'^\mu k'^\nu -\eta^{\mu\nu}k'^2\right] k'^2.
\end{equation}
This term is the higher-derivative counterpart of the term with two derivatives in Eq. (\ref{PIMUNUK2}) and turns out to be wonderfully gauge invariant, as it should.

\subsection{The polarization Tensor up to $\mathcal{O}(k'^4)$}
Summarizing our results, we write the polarization tensor as
$\Pi^{\mu\nu}(k')=\Pi^{\mu\nu}_{even}+\Pi^{\mu\nu}_{odd}$. The even and odd parts are 
\begin{eqnarray}
\Pi^{\mu\nu}_{even}&=&\Pi^{\mu\nu}_{2k'} + \Pi^{\mu\nu}_{4k'} = 
 (k'^\mu k'^\nu-\eta^{\mu\nu}k'^2)\left( 1 +\frac{1}{10} \frac{k'^2}{m^2}\right) \frac{e^2}{12\pi|m|} \equiv
(k'^\mu k'^\nu-\eta^{\mu\nu}k'^2) \, \pi_0(k'), \\ 
\Pi^{\mu\nu}_{odd}&=&\Pi^{\mu\nu}_{1k´}+\Pi^{\mu\nu}_{3k'} 
= -i\epsilon^{\mu\beta\nu} \, k'_\beta \left(1+\frac{1}{12}\frac{k'^2}{m^2}\right) \frac{e^2}{4\pi}\mathrm{Sgn}(m) \equiv -i\epsilon^{\mu\nu\beta}\, k_\beta \, \pi_{\rm CS}(k'),
\end{eqnarray}
respectively, which defines $\pi_0(k')$ and $\pi_{\rm CS}(k')$. 
The final result is
\begin{equation}
	\Pi^{\mu\nu}(k')=(k'^\mu k'^\nu-\eta^{\mu\nu}k'^2) \pi_0 (k´)-i\epsilon^{\mu\beta\nu}\, k'_\beta\pi_{\rm CS}(k'),
\end{equation}
with
\begin{equation}
	\pi_0(k')=\frac{e^2}{12\pi|m|} \, \left( 1 +\frac{1}{10} \frac{k'^2}{m^2}\right),  \qquad  \pi_{\rm CS}(k')= \frac{e^2}{4\pi}\mathrm{Sgn}(m) \, 
 \left(1+\frac{1}{12}\frac{k'^2}{m^2}\right). 
 \label{FINALPIS}
\end{equation}
It is easy to see that the full polarization tensor is gauge-invariant, i.e.,
\begin{equation}
	k'^\mu \Pi_{\mu\nu}(k') = k'^\nu \Pi_{\mu\nu}(k')=0.
\end{equation}
As a final consistency check of our calculations, we start from the non-perturbative  result of the complete polarization tensor (see for example Ref. \cite{fradkinrev})
\begin{equation}
\pi_0(k')=-\frac{|m|}{4\pi k'^2}+\frac{1}{8\pi \sqrt{k'^2}}\left(\frac{4m^2}{k'^2}+1\right)\sinh^{-1}\left(\frac{\sqrt{k'^2}}{\sqrt{4m^2-k'^2}}\right), \qquad 
\pi_{CS}(k')=\frac{m}{2\pi\sqrt{k'^2}}\sinh^{-1}\left(\frac{\sqrt{k'^2}}{\sqrt{4m^2-k'^2}}\right),
\end{equation}
and perform a power series expansion in the low-energy regime $|k'^2|<<m^2$, retaining terms up to order $k'^2/m^2$. We recover exactly the expressions shown in Eq. (\ref{FINALPIS}).

Finally, we go back to the expression for the effective action in coordinate space. Since we have two powers of $A_\beta$, and only one type of derivative ($k'_\alpha=i\partial'_\alpha $), it is a simple exercise to recover the final expression for the local effective action. The result is 
\beq
S_{\rm eff}^{(2)}= \frac{e^2}{2} \int d^4 x \, A^\nu(x) \Big[ \frac{1}{12 \pi m} \Big(1- \frac{1}{10 }\frac{\partial^2}{m^2} \Big) \,\Big( -\partial_\mu \partial_\nu + \eta_{\mu\nu}\partial^2\Big) + \frac{\mathrm{sgn}(m)}{4\pi} \epsilon_\nu{}^{\mu\beta } \, \partial_\beta \Big(1-\frac{1}{12} \frac{\partial^2}{m^2}\Big)\Big] A_\mu(x),
\label{FULLACT}
\eeq
where all the derivatives are acting on $A_\mu(x)$. 

\section{Summary and Discussion}

With an eye on exploring alternative methods to calculate effective electromagnetic Lagrangians in Dirac-like models, we have used an early version of the derivative expansion  method  \cite{Novikov84,Zuk85,Fraser85,Aitchison1985}, focusing on obtaining the corresponding lowest-order derivative corrections. This approach directly provides local effective Lagrangians and allows selecting from the outset both the powers of the electromagnetic potential to be considered, together with the number of relevant derivatives. We highlight these features as an advantage over the non-local effective actions resulting from the standard calculation of the vacuum polarization tensor, for example. 

For the benefit of the reader, a brief description of the method was presented along  with a summary of how the general effective electromagnetic action is obtained.  The first case considered is $(3+1)$ massless electrodynamics with an additional coupling $b_\mu \gamma^5 \gamma^\mu$, which is relevant to the description of the simplest Weyl semimetal with two Dirac cones separated in momentum by $\mbf{b}$ and in energy by $b_0$, in the Brillouin zone. \cite{CMHEFF}. Restricting ourselves to the axial contribution of the vacuum polarization tensor, we recovered   Carroll-Field-Jackiw  (CFJ)  electrodynamics in the zeroth order of the derivative expansion. As is well known, the numerical coefficient of this action is finite but undetermined, which is regularization dependent. Choosing alternative regularizations, we match our result with some in the literature, as shown in Table \ref{tab1}. The lowest-order derivative correction is calculated for the first time in a massless theory, and the coefficient is again finite but regularization dependent. The calculation in spherical coordinates yields
\beq
J_{\rm CFJ}^\mu=\frac{e^2}{48 \pi^2 b^2}\, \epsilon^{\mu \alpha\beta \gamma} \, b_\alpha \partial^2 F_{\beta\gamma},
\eeq
which constitutes an addition to the CFJ equations.  Notice that for $b^\mu=(0, \mbf{b})$ the above current is a covariant generalization ($\boldsymbol{\nabla^2} \rightarrow \partial^2$) of the additional terms in the Maxwell equations proposed in Ref. \cite{BARREDO} when $\psi(x) \sim \mbf{b}\cdot \mbf{x}$. 

The second case we dealt with is $(2+1)$ massive  ($m$) electrodynamics, which can be used as a model to describe the plane interface between two topological insulators, for example. The full effective action up to fourth order in the derivatives is given by Eq. (\ref{FULLACT}) and the resulting current is given by
\beq
J_{(2+1)}^\mu= \frac{e^2}{12 \pi} \frac{1}{m} \Big( 1-\frac{1}{10} \frac{\partial^2}{m^2}\Big) \partial_\alpha F^{\alpha \mu}
-\frac{e^2}{8\pi} {\rm sgn}(m) \Big( 1-\frac{1}{12} \frac{\partial^2}{m^2} \Big) \epsilon^{\mu \alpha\beta} F_{\alpha \beta}.
\label{2+1CURR}
\eeq
Let us observe that, in our case, we do not have a superimposed magnetic field.
From the above equation, the relativistic transverse conductivity resulting from the Hall current (the contribution proportional to $\epsilon^{\mu \alpha\beta}$) is
\beq
\sigma_{xy}(k)= {\rm sgn}(m) \frac{e^2}{(4\pi)}\Big( 1+ \frac{1}{12}\frac{k^2}{m^2}\Big).
\label{SIGMAXYT}
\eeq
The first term,  $\sigma_{xy}(k=0)$, corresponds to the usual Chern-Simons contribution to the transverse conductivity.  When reintroducing units, the prefactor in Eq. (\ref{SIGMAXYT}) simplifies to ${\rm sgn}(m) e^2/(4\pi \hbar) = {\rm sgn}(m) e^2/(2h)$. The second term provides a relativistic generalization of the $\xi / \kappa$-term introduced in Ref. \cite{TVM}, in the ratio $\sigma_{xy}(k)/ \sigma_{xy}(k=0)$, for the case of a Galilean invariant theory. In the latter case, the ratio $\xi / \kappa $ is related to the Hall viscosity $\eta^H$. In single-component Galilean-invariant fluids and solids with particle number symmetry,
the momentum density is proportional to the particle number current, resulting in a Ward identity that links the viscosity and conductivity tensors \cite{HOYOS1,HOYOS2,BRADLYN,HVISC3,37,40}.

The relativistic case in a Chern insulator modeled as a free massive Dirac fermion has been considered in Refs. \cite{HUGHES1,HUGHES2} from the perspective of the response of the material due to strain deformations induced by a curved metric. In particular, the Hall viscosity appears in the corresponding torsional response.  Up to our knowledge, the possible relation between the Hall conductivity and the Hall viscosity remains an open problem in the case of relativistic systems. 

The lesson we learn is that these naturally emerging correction terms in the effective action should be incorporated into the phenomenological proposals for higher-order derivative corrections to assess the relative significance of all the terms included up to a given order. 

 Since the scope of this work is more on a technical description of the derivative expansion method we have not considered specific  physical applications  of the corrections we have found in some effective electromagnetic actions. However, in the Introduction we have mentioned important applications already developed or envisaged  in the literature. This is particularly noteworthy in the case of the anomalous Hall conductivity $\sigma_{xy}$ arising  at the interface between two 3D topological insulators, and resulting from  a $2+1$ QED at the boundary. As mentioned in the introduction, higher order derivative corrections  are crucial  for an alternative simpler way to measure the Hall viscosity,  an additional topological property of the quantum states at the boundary, for a particular class of materials. Additionally, as shown in Ref. \cite{MECHELEN2}, the momentum dependence of the optical parameters (particularly  of $\sigma_{xy}$), obtained through higher-order derivative corrections, for example, is mandatory to explain the existence of topological bosonic phases, in particular the photonic ones. This is a topic much less studied in the literature in comparison with the fermionic case. In a forthcoming publication \cite{RAU} we will consider the effect of the derivative corrections on $\sigma_{xy}$ upon two physical effects: (i) the Kerr and Faraday rotation angles induced when an electromagnetic wave impinges on a topological insulator and (ii) corrections to the resulting magnetoelectric effect when a point charge is placed near a topological insulator,  which is related to the generation of an image magnetic charge.

\acknowledgments

 R.M.vD. and L.F.U. acknowledge support from the projects
CONACyT (M\'exico) \# CF-428214 and DGAPA-UNAM \# AG100224 . R.M.vD. was supported by the DGAPA-UNAM Posdoctoral Program .

\appendix 

\section{Integrals in Four-Dimensional Spherical Coordinates}\label{spherical}

In this appendix, we  explicitly calculate the integrals appearing in equations (\ref{CFJNI}) and (\ref{CFJHDNI}). To begin with, let us  compute the following integral
\begin{equation}\label{II}
	I^{(n)}=\int \frac{d^4p}{\left(2\pi\right)^4}\frac{1}{\left[\left(p+\chi b\right)^2\right]^n}=\int \frac{d^4p}{\left(2\pi\right)^4}\frac{1}{\left(p^2+b^2+2\chi p_\mu b^\mu \right)^n}.
\end{equation}
We perform a Wick rotation to move into Euclidean space,  choosing  
\begin{eqnarray}
	p_0 &\,\, \rightarrow&  \,\, ip_0 \qquad \implies  \qquad d^4p\rightarrow id^4p \hspace{10mm}\mathrm{and }\hspace{10mm}p^2\rightarrow -p^2, \\
	b_0 & \,\, \rightarrow& \,\, ib_0  \qquad \implies \qquad b^2\rightarrow-b^2\hspace{10mm}\mathrm{and}\hspace{10mm}p_\mu b^\mu\rightarrow-p_\mu b^\mu,
\end{eqnarray}
which change $I^{(n)}$ into 
\begin{equation}
	I^{(n)}_E=(-1)^n i\int \frac{d^4p}{\left(2\pi\right)^4}\frac{1}{\left(p^2+b^2+ 2 \chi p_\mu b^\mu \right)^n},
\end{equation}
where all the scalar products are now in the Euclidean metric $\delta^{\mu\nu}= {\rm diag}(1,1,1,1)$.
We  solve this integral in 4-dimensional spherical coordinates, where the differential volume element is 
\begin{equation}
	d^4p \,\,  \rightarrow  \,\, dp \, d\theta_1 \,  d\theta_2 \, d\theta_3 \,  p^3  \,\sin^2\theta_1 \, \sin\theta_2,
\end{equation}
with the ranges
\begin{equation}
	0 \leq p < \infty \hspace{6mm},\hspace{6mm} 0 \leq \theta_1 \leq \pi \hspace{6mm},\hspace{6mm}0 \leq \theta_2 \leq \pi\hspace{6mm}y\hspace{6mm}0 \leq \theta_3 \leq 2\pi.
\end{equation}
In this coordinate system, $p$ represents the radial distance from the origin in four-dimensional Euclidean space, $\theta_1$ is the first polar angle, measuring the inclination of $p_\mu$ with respect to the fourth coordinate. $\theta_2$ is the second polar angle, describing the orientation of the vector within the three-dimensional subspace perpendicular to the fourth coordinate. Finally, $\theta_3$ is the azimuthal angle, representing rotation around the remaining axis in the three-dimensional subspace. Without loss of generality, we fix the coordinate system so that the four-vector $b_\mu$
  is aligned along the fourth coordinate. This choice simplifies the calculation of the integral, as $b_\mu$
  will then form a well-defined angle $\theta_1$ with the radial four-vector $p_\mu$, thereby reducing the integral expression in terms of this angle.
Hence, the integral (\ref{II}) takes the form
\begin{equation}
	I^{(n)}_E= \frac{(-1)^ni}{\left(2\pi\right)^4}\int_{0}^{\infty}dp p^3 \int_{0}^{\pi}d\theta_1 \, \sin^2\theta_1\underbrace{\int_{0}^{\pi}d\theta_2  \, \sin\theta_2\int_{0}^{2\pi}d\theta_3}_{=4\pi}\frac{1}{\left(p^2+b^2+2\chi \,p \, b \, \mathrm{cos}\theta_1 \right)^n},
\end{equation}
\begin{equation} \label{IE}
	I^{(n)}_E= \frac{(-1)^ni}{4\pi^3}\int_{0}^{\infty}dp p^3 \int_{0}^{\pi}d\theta_1\frac{\sin^2\theta_1}{\left(p^2+b^2+2\chi \, p \, b \, \mathrm{cos}\theta_1 \right)^n}.
\end{equation}
The general result for the integral over $d\theta_1$ can be found in Ref. \cite{grad}, and  it is 
\begin{equation}
	\int_{0}^{\pi}\frac{\sin^{2\mu-1}(x)}{\left(1+2a \cos(x)+a^2\right)^\nu}dx=B\left(\mu,\frac{1}{2}\right)F\left(\nu,\nu-\mu+\frac{1}{2};\mu+\frac{1}{2};a^2\right),\qquad \mathrm{Re}(\mu) >0\, ,\qquad  |a|<1,
 \label{RESULT}
\end{equation}
where $B\left(\mu,\frac{1}{2}\right)$ is the beta function, defined as 
\begin{equation}
	B\left(\alpha, \gamma\right)=\frac{\Gamma\left(\alpha\right)\Gamma\left(\gamma\right)}{\Gamma\left(\alpha+\gamma \right)},
\end{equation}
and $F\left(\nu,\nu-\mu+\frac{1}{2};\mu+\frac{1}{2};a^2\right)$ is the hypergeometric function, given by 
\begin{equation}
	F\left( a,b;c;z \right)=\sum_{n=0}^{\infty}\frac{(a)_n(b)_n}{(c)_n}\frac{z^n}{n!}, \qquad  (a)_n=\frac{\Gamma\left(a+n\right)}{\Gamma\left(a\right)}.
\end{equation}
To apply the result (\ref{RESULT}) to our integral (\ref{IE}), we  split it into two parts
\begin{equation}
	I_E^{(n)}=I_E^{(n)p<b}+I_E^{(n)b<p},
\end{equation}
such that we can separately deal with the restriction $|a| < 1$.

For the first case $p< b$, we choose   $a= \chi \frac{p}{b}$, which satisfies $a^2 < 1 $,  together with   $\mu=\frac{3}{2}$, $\nu=n$, and obtain
\begin{equation}
		I_E^{(n)\,p<b}= \frac{(-1)^ni}{4\pi^3b^{2n}}\int_{0}^{b-\epsilon}dp p^3 \int_{0}^{\pi}d\theta_1\frac{\sin^2\theta_1}{\left(1+2a \mathrm{cos}\theta_1 +a^2\right)^n}.
	\end{equation}
 We set the upper limit of the $p$-integration as $b-\epsilon$, where $\epsilon \rightarrow 0^+$ is introduced to regularize a divergence that arises at $p=b$ after the angular integration. Performing the angular integration according to (\ref{RESULT}) with the result
	\begin{eqnarray}
		I_E^{(n)\,p < b}&=&\frac{(-1)^n i}{4\pi^3b^{2n}}\int_{0}^{b-\epsilon}dp p^3 B\left(\frac{3}{2},\frac{1}{2}\right)F\left(n,n-1;2;\frac{p^2}{b^2}\right) = \frac{(-1)^n i}{8\pi^2b^{2n}}\int_{0}^{b-\epsilon}dp p^3 F\left(n,n-1;2;\frac{p^2}{b^2}\right),
	\end{eqnarray}
after substituting   $B\left(\frac{3}{2},\frac{1}{2}\right)=\frac{\pi}{2}$.

Similarly, when $p> b$,  we choose $a= \chi\frac{b}{p}$  which satisfies  $a^2 < 1$, together with $\mu=\frac{3}{2}$, $\nu=n$. This yields
\begin{equation}
		I_E^{(n)\,p>b}=\frac{(-1)^ni}{4\pi^3}\int_{b+\epsilon}^{\Lambda}dp p^{3-2n} \int_{0}^{\pi}d\theta_1\frac{\sin^2\theta_1}{\left(1+2a \mathrm{cos}\theta_1 +a^2\right)^n}.
	\end{equation}
Using again (\ref{RESULT}) for the angular integration,  we find
	\begin{eqnarray}
		\hspace{-0.5cm} I_E^{(n)\,p>b}&=& \frac{(-1)^ni}{4\pi^3}\int_{b+\epsilon}^{\Lambda}dp p^{3-2n}B\left(\frac{3}{2},\frac{1}{2}\right)F\left(n,n-1;2;\frac{b^2}{p^2}\right) =
		\frac{(-1)^ni}{8\pi^2}\int_{b+\epsilon}^{\Lambda}dp p^{3-2n}F\left(n,n-1;2;\frac{b^2}{p^2}\right).
	\end{eqnarray}
 Here, we introduced a the cutoff $\Lambda$ to regularize the integral as $p\rightarrow \infty$.
 
 Summarizing, the integral $I_E^{(n)}$ results
\begin{equation}
	I^{(n)}_E=\frac{(-1)^n i}{8\pi^2b^{2n}}\int_{0}^{b-\epsilon}dp p^3 F\left(n,n-1;2;\frac{p^2}{b^2}\right)+\frac{(-1)^ni}{8\pi^2}\int_{b+\epsilon}^{\Lambda}dp p^{3-2n}F\left(n,n-1;2;\frac{b^2}{p^2}\right),
\end{equation}
from which we can further compute the required  integrals for $n=1$ and $n=2$.
\begin{itemize}
	\item Case $n=1$
	\begin{equation}
		I^{(1)}_E=-\frac{i}{8\pi^2b^{2}}\int_{0}^{b-\epsilon}dp p^3 F\left(1,0\, ; 2\, ;\frac{p^2}{b^2}\right)-\frac{i}{8\pi^2}\int_{b+\epsilon}^{\Lambda}dp p F\left(1,0 \, ; 2\, ;\frac{b^2}{p^2}\right),
	\end{equation}
where $F\left(1,0\, ;2\, ;\frac{p^2}{b^2}\right)=F\left(1,0 \, ;2\, ;\frac{b^2}{p^2}\right)=1$, supplying
	\begin{eqnarray}
		I^{(1)}_E&=&-\frac{i}{8\pi^2b^{2}}\left. \frac{p^4}{4} \right|^{b-\epsilon}_{0} -\frac{i}{8\pi^2}\left.\frac{p^2}{2}\right|_{b+\epsilon}^{\Lambda} = \frac{ib^2}{32\pi^2} -\frac{i\Lambda}{16\pi^2}.
	\end{eqnarray}
Returning to Minkowski spacetime, we have
	\begin{equation}
I^{(1)}=\int \frac{d^4p}{\left(2\pi\right)^4}\frac{1}{\left(p+\chi b \right)^2}	=-\frac{ib^2}{32\pi^2}-\frac{i\Lambda}{16\pi^2}. 
	\end{equation}
\item Case $n=2$
	\begin{equation}
		I^{(2)}_E=\frac{i}{8\pi^2b^4}\int_{0}^{b}dp p^3 F\left(2,1\, ;2\, ;\frac{p^2}{b^2}\right)+\frac{i}{8\pi^2}\int_{b}^{\Lambda}dp p^{-1}F\left(2,1\, ;2 \,;\frac{b^2}{p^2}\right),
	\end{equation}
where $F\left(2,1 \, ;2\, ;\frac{p^2}{b^2}\right)=\frac{b^2}{b^2-p^2}$ and $F\left(2,1\, ;2 \, ;\frac{b^2}{p^2}\right)=\frac{p^2}{p^2-b^2}$. Then we obtain 
	\begin{equation}
		I^{(2)}_E=\frac{i}{8\pi^2}\left[ \frac{1}{b^2}\int_{0}^{b-\epsilon}dp \frac{p^3}{b^2-p^2}+\int_{b+\epsilon}^{\Lambda}dp \frac{p}{p^2-b^2}\right].
	\end{equation}
Making the substitution $p^2=x \implies dx=2pdp$, we have
	\begin{equation}
		I^{(2)}_E=\frac{i}{16\pi^2}\left[ \frac{1}{b^2}\int_{0}^{b^2(1-\epsilon)}dx \frac{x}{b^2-x}+\int_{b^2(1+\epsilon)}^{\Lambda^2}dx \frac{1}{x-b^2}\right].
	\end{equation}
Integrating over $x$ supplies
	\begin{eqnarray}
		I^{(2)}_E&=&\frac{i}{16\pi^2}\left[ \left.\left[ -\mathrm{ln}\left(x-b^2\right)-\frac{x}{b^2}\right]\right|_0^{b^2(1-\epsilon)}+ \left.\left[\mathrm{ln}(x-b^2)\right]\right|_{b^2(1+\epsilon)}^{\Lambda^2} \right]
 = \frac{i}{16\pi^2}\left[-1+\mathrm{ln}\left(\frac{\Lambda^2-b^2}{\epsilon^2b^2}\right)\right]. 
	\end{eqnarray}
Finally, in Minkowski space, we have 
	\begin{equation}\label{I2}
			I^{(2)}=\int \frac{d^4p}{\left(2\pi\right)^4}\frac{1}{\left[ \left(p+ \chi b\right)^2\right]^2}=\frac{i}{16\pi^2}\left[-1+\mathrm{ln}\left(\frac{-\Lambda^2-b^2}{\epsilon^2b^2}\right)\right].  
	\end{equation}
\end{itemize}
Let us remark that the integrals $I^{(1)}$ and  $I^{(2)}$  are independent of $\chi$, as it enters in the combination $(\chi b)^2=b^2$. Starting from  these integrals, we next calculate the different combinations of the type 
\beq
    I^{(n)}_{\alpha\beta....}= \int \frac{d^4 p}{(2 pi)^4} \frac{\pi_\alpha \, \pi_\beta \, ...}{\pi^{2n}},
    \eeq
which are required in section \ref{IIIB}. We obtain these integrals by taking convenient derivatives with respect to $b^\mu$ in the results for $I^{(1)}$ and  $I^{(2)}$. We recall that $\pi_\alpha= (p+\chi b)_\alpha$ with $\pi^{2n}= (\pi_\alpha \pi^\alpha)^n$. The results are 
\beq
I_{ \alpha}^{(2)}= \int \frac{d^4p}{\left(2\pi\right)^4}\frac{\left(p+\chi b\right)_\alpha}{\left[\left(p+\chi b\right)^2\right]^2}    =-\frac{\chi}{2}\frac{\partial  I^{(1)}}{\partial b^\alpha}= \frac{i\chi b_\alpha}{32\pi^2},
\eeq
\begin{equation}
	I_\alpha^{(3)}=\int \frac{d^4p}{\left(2\pi\right)^4}\frac{\left(p+\chi b\right)_\alpha}{\left[\left(p+\chi b\right)^2\right]^3}=-\frac{\chi}{4} \frac{\partial  I^{(2)}}{\partial b^\alpha} =\frac{i\chi b_\alpha}{32\pi^2b^2},
\end{equation}
Next, we  calculate the derivative of $I^{(2)}_\alpha$. Differentiating both sides, we have
\begin{eqnarray}
\frac{\partial I_\alpha^{(2)}}{\partial b^\beta}&=&\frac{\partial}{\partial b^\beta}\left[ \frac{i\chi b_\alpha}{32\pi^2} \right]
,\nonumber \\
	\chi  \, \eta_{\alpha \beta }  I^{(2)}-4\chi 
	I^{(3)}_{\alpha \beta}&=&\frac{i\chi }{32\pi^2} \eta_{\alpha\beta},
\end{eqnarray}
which  leads to 
\begin{eqnarray}
I^{(3)}_{\alpha\beta}=	\int \frac{d^4p}{\left(2\pi\right)^4}\frac{\left(p+\chi b\right)_\alpha\left(p+\chi b\right)_\beta}{\left[\left(p+\chi b\right)^2\right]^3}=\eta_{\alpha\beta } \Big(\frac{1}{4}  I^{(2)} -\frac{i}{128\pi^2}\Big)=
i\eta_{\alpha\beta} \frac{1}{64\pi^2} \Big( \mathrm{ln}\left(\frac{-\Lambda^2-b^2}{\epsilon^2b^2}\right)-\frac{3}{2} \Big).
\end{eqnarray}
Our last integral is obtaining by differentiating $I_{\alpha\beta}^{(3)}$ and following  analogous steps as in the previous cases. We obtain 
\begin{equation}\label{I4abc}
		I^{(4)}_{\alpha\beta\lambda}=\int \frac{d^4p}{\left(2\pi\right)^4}\frac{\left(p+\chi b\right)_\alpha \left(p+\chi b\right)_\beta \left(p+\chi b\right)_\lambda}{\left[\left(p+\chi b\right)^2\right]^4}= \frac{i \chi}{192 \pi^2 b^2}\, ( \eta_{\beta\lambda}b_\alpha+ \eta_{\alpha\lambda}b_\beta+ \eta_{\alpha\beta}b_\lambda).
\end{equation}

\section{Symmetrization of Integrals in Momentum Space}
 When integrating products of  a function of the invariant $p^2$ times tensors formed by products of  the momentum $p^\mu$, the use of Lorentz covariance allows simplifying the expressions by replacing such tensors with combinations of the metric $\eta^{\alpha\beta}$ times powers of the scalar $p^2$. In the $2+1$-dimensional case considered in this work we identify the  following useful simplifications: 
\begin{equation}\label{Ip}
	\int\frac{d^3p}{(2\pi)^3}f(p^2)\underbrace{p^\mu p^\nu p^\rho \dots}_{\rm odd \, \, number}=0,
\end{equation}
\begin{equation}\label{Ipp}
	\int\frac{d^3p}{(2\pi)^3}f(p^2)p^\mu p^\nu=\frac{\eta^{\mu\nu}}{3}\int\frac{d^3p}{(2\pi)^3}f(p^2)p^2,
\end{equation}
\begin{equation}\label{Ipppp}
	\int\frac{d^3p}{(2\pi)^3}f(p^2)p^\mu p^\nu p^\alpha p^\beta=\frac{\eta^{\mu\nu}\eta^{\alpha\beta}+\eta^{\mu\alpha}\eta^{\nu\beta}+\eta^{\mu\beta}\eta^{\nu\alpha}}{15}\int\frac{d^3p}{(2\pi)^3}f(p^2)p^4,
\end{equation}
\begin{eqnarray}\label{I6p}
	\int\frac{d^3p}{(2\pi)^3}f(p^2)p^\mu p^\nu p^\alpha p^\beta p^\lambda p^\rho &=&\frac{1}{105} \left(\eta^{\mu\nu}(\eta^{\alpha\beta}\eta^{\lambda\rho}+\eta^{\alpha\lambda}\eta^{\beta\rho}+\eta^{\alpha\rho}\eta^{\beta\lambda})+\eta^{\mu\alpha}(\eta^{\nu\beta}\eta^{\lambda\rho}+\eta^{\nu\lambda}\eta^{\beta\rho}+\eta^{\nu\rho}\eta^{\beta\lambda})\right. \nonumber \\ && \left. + \eta^{\mu\beta}(\eta^{\nu\alpha}\eta^{\lambda\rho}+\eta^{\nu\lambda}\eta^{\alpha\rho}+\eta^{\nu\rho}\eta^{\alpha\lambda})+\eta^{\mu\lambda}(\eta^{\nu\alpha}\eta^{\beta\rho}+\eta^{\nu\beta}\eta^{\alpha\rho}+\eta^{\nu\rho}\eta^{\alpha\beta})\right. \nonumber \\ && \left. + \eta^{\mu\rho}(\eta^{\nu\alpha}\eta^{\beta\lambda}+\eta^{\nu\beta}\eta^{\alpha\lambda}+\eta^{\nu\lambda}\eta^{\alpha\beta})\right)\int\frac{d^3p}{(2\pi)^3}f(p^2)p^6.
\end{eqnarray}
\section{The Master Integral}\label{APMI}
We solve the integrals over $p$ which are required in section \ref{IIIC}, when dealing  with the calculation of the effective action in $2+1$  QED. The master integral is 
\begin{equation}
	I^\beta_\alpha=\int\frac{d^3p}{(2\pi)^3} \frac{p^{2\beta}}{\left(p^2-m^2\right)^\alpha}.
\end{equation}
To solve this integral, we first go to Euclidean space through a Wick rotation given by
\begin{equation}
	p_0 \,\,  \rightarrow    \,\, ip_0 \quad  \implies  \quad p^2 \,\, \rightarrow  \,\, -p^2,
\end{equation}
such that the integral takes the form
\begin{equation}
	I^\beta_\alpha=(-1)^{\alpha+\beta}i\int\frac{d^3p}{(2\pi)^3} \frac{p^{2\beta}}{\left(p^2+m^2\right)^\alpha},
  \end{equation}
where now the scalar products are in the Euclidean metric ${\rm diag}=(1,1,1,1)$.  
Passing to spherical coordinates, we have
\begin{equation}
	I^\beta_\alpha=i\frac{(-1)^{\alpha+\beta}}{(2\pi)^3}\int_{0}^{2\pi}d\phi \int_{0}^{\pi}d\theta \sin\theta \int_{0}^{\infty}dp \, p^2 \frac{p^{2\beta}}{\left(p^2+m^2\right)^\alpha}.
\end{equation}
Given the spherical symmetry of the integrand, we can directly integrate over the angular variables, obtaining
\begin{equation}
	I^\beta_\alpha=i\frac{(-1)^{\alpha+\beta}}{2\pi^2}\int_{0}^{\infty}dp  \frac{p^{2(\beta+1)}}{\left(p^2+m^2\right)^\alpha}.
\end{equation}
Now, we perform the  change of variables $p^2=xm^2 \,\, \implies \,\, dp=\frac{|m|x^{-1/2}}{2}dx$, therefore
\begin{eqnarray}
I^\beta_\alpha&=&i\frac{(-1)^{\alpha+\beta}}{4\pi^2}\int_{0}^{\infty}dx x^{-1/2} |m|  x^{\beta+1}|m|^{2(\beta+1)}\left(x|m|^2+|m|^2\right)^{-\alpha}, \nonumber \\
	&=& i\frac{(-1)^{\alpha+\beta}|m|^{-2\alpha+2\beta+3}}{4\pi^2}\int_0^{\infty} dx \, x^{\beta+\frac{3}{2}-1} \, (x+1)^{-\alpha}.
\end{eqnarray}
The integral in $x$ is a standard Beta function
\begin{equation}
	B(\gamma,\lambda)=\int_{0}^{\infty}dx \, x^{\gamma-1}(x+1)^{-\gamma-\lambda}=\frac{\Gamma(\gamma)\Gamma(\lambda)}{\Gamma(\gamma+\lambda)},
\end{equation}
which leads to the  final result
\begin{equation}\label{IM}
	I^\beta_\alpha=i\frac{(-1)^{\alpha +\beta }  |m|^{-2 \alpha +2 \beta +3}}{4 \pi^2 } B\left(\beta +\frac{3}{2},\alpha -\beta -\frac{3}{2}\right).
\end{equation}
Now, we evaluate this integral for the different values of the  parameters $\alpha$ and $\beta$ that occur in section \ref{IIIC}, obtaining
\begin{equation}
	I^0_1=\frac{i|m|}{4\pi}
\hspace{5mm},\hspace{5mm}
	I^1_2=\frac{3i|m|}{8\pi}
\hspace{5mm},\hspace{5mm}
	I^0_2=\frac{i}{8\pi|m|},
\end{equation}
\begin{equation}
	I^1_3=\frac{3i}{32\pi|m|}\hspace{5mm},\hspace{5mm}
	I^2_4=\frac{5i}{64\pi|m|}\hspace{5mm},\hspace{5mm}
	I^0_3=-\frac{i}{32\pi|m|^3},
\end{equation}
\begin{equation}
	I^1_4=-\frac{i}{64\pi|m|^3},
\hspace{5mm},\hspace{5mm}
	I^2_5=-\frac{5i}{512\pi|m|^3}\hspace{5mm},\hspace{5mm}
	I^3_6=-\frac{7i}{1024\pi|m|^3}.
\end{equation}

\end{document}